\documentclass[12pt]{iopart}
\usepackage{iopams}
\expandafter\let\csname equation*\endcsname\relax
\expandafter\let\csname endequation*\endcsname\relax
\usepackage{amsmath}
\pdfoutput=1
\usepackage{graphicx,psfrag,color}
\usepackage[utf8]{inputenc}
\usepackage[T1]{fontenc}
\usepackage{lmodern}
\usepackage{times}
\usepackage{mathptmx}
\usepackage{epstopdf}
\usepackage{xcolor}
\usepackage[colorlinks,linkcolor=blue,urlcolor=blue,citecolor=blue]{hyperref}
\usepackage[normalem]{ulem}
\usepackage{xspace}
\usepackage{float}
\usepackage{bm}
\usepackage{cite}
\usepackage{lineno}
\hypersetup{
    colorlinks=true,
    linkcolor=blue,
    filecolor=magenta,
    urlcolor=cyan,
    citecolor=cyan
}

\def\la{\langle}
\def\ra{\rangle}

\DeclareSymbolFont{bbold}{U}{bbold}{m}{n}
\DeclareSymbolFontAlphabet{\mathbbold}{bbold}
\makeatletter
\newcommand*\bigcdot{\mathpalette\bigcdot@{.55}}
\newcommand*\bigcdot@[2]{\mathbin{\vcenter{\hbox{\scalebox{#2}{$\m@th#1\bullet$}}}}}
\makeatother

\def\im{\mathrm{i}}
\def\ee{\mathrm{e}}
\def\bra#1{\mathinner{\langle{#1}|}}
\def\ket#1{\mathinner{|{#1}\rangle}}
\providecommand{\keywords}[1]{\textbf{\textit{Keywords:}} #1}

\begin{document}

\newcommand{\gae}{\lower 2pt \hbox{$\, \buildrel {\scriptstyle >}\over {\scriptstyle
\sim}\,$}}
\newcommand{\lae}{\lower 2pt \hbox{$\, \buildrel {\scriptstyle <}\over {\scriptstyle
\sim}\,$}}

\title{Quantum unitary evolution interspersed with repeated non-unitary interactions at random times: The method of stochastic Liouville equation,  and two examples of interactions in the context of a tight-binding chain}
\author{Debraj Das$^1$, Sushanta Dattagupta$^2$, and
Shamik Gupta$^{3,4}$}
\address{$^1$Department of Engineering Mathematics, University of
Bristol, Bristol BS8 1TW, United Kingdom\\ $^2$Bose Institute, Kolkata
700054, India \\
$^3$Department of Physics, Ramakrishna Mission Vivekananda
Educational and Research Institute, Belur Math, Howrah 711202, India\\$^4$Quantitative Life Sciences Section, ICTP–Abdus Salam International Centre for Theoretical Physics, Strada Costiera 11, 34151 Trieste, Italy}
\ead{debraj.das@bristol.ac.uk,sushantad@gmail.com,shamikg1@gmail.com}
\date{\today}
\begin{abstract}  
In the context of unitary evolution of a generic quantum system interrupted at random times with non-unitary evolution due to interactions with either the external environment or a measuring apparatus,  we adduce a general theoretical framework to obtain the average density operator of the system at any time during the dynamical evolution. The average is with respect to the classical randomness associated with the random time intervals between successive interactions,  which we consider to be independent and identically-distributed random variables. The formalism is very general in that it applies to any quantum system,  to any form of non-unitary interaction, and to any probability distribution for the random times.  We provide two explicit applications of the formalism in the context of the so-called tight-binding model relevant in various contexts in solid-state physics, e.g., in modelling nano wires.   Considering the case of one dimension,  the corresponding tight-binding chain models the motion of a charged particle between the sites of a lattice,  wherein the particle is for most times localized on the sites, but which owing to spontaneous quantum fluctuations tunnels between nearest-neighbour sites.   We consider two representative forms of interactions,  one that implements stochastic reset of quantum dynamics in which the density operator is at random times reset to its initial form, and one in which projective measurements are performed on the system at random times.  In the former case,  we demonstrate with our exact results how the particle is localized on the sites at long times, leading to a time-independent mean-squared displacement (MSD) of the particle about its initial location. This stands in stark contrast to the behavior in the absence of interactions,  when the particle has an unbounded growth of the MSD in time, with no signatures of localization.  In the case of projective measurements at random times,  we show that repeated projection to the initial state of the particle results in an  effective suppression of the temporal decay in the probability of the particle to be found on the initial state.  The amount of suppression is comparable to the one in conventional Zeno effect scenarios,  but which however does not require performing measurements at exactly regular intervals that are hallmarks of such scenarios. 
\end{abstract}
\date{\today}
\keywords{Stochastic particle dynamics, Stationary states, Quantum wires}
\maketitle
\tableofcontents
\section{Introduction}
\label{secI}

Quantum dynamics governed by a Hermitian Hamiltonian is unitary, leading to coherent evolution of the wave function, which is essential for preserving quantum information. Most real systems however are dissipative, causing loss of phase coherence. In contemporary times, with the laboratory advances in the fabrication of nano-mesoscopic systems, there have been renewed hopes of realizing coherent processes. But the flip side is that a small device is inevitably under non-negligible interaction with its environment, making an otherwise unitary evolution non-unitary. The challenge therefore is to carefully analyse the influence of dissipative couplings and devise schemes to minimize such an influence.

Dissipation manifests itself in two distinct ways. One,  wherein the subsystem of interest is part of an `open system,' in which the environment acts as a heat bath with which the subsystem is in continual exchange of energy. Such studies fall in the realm of dissipative quantum systems~\cite{Weiss:2012,Dattagupta:2004}. The second scenario is one in which one manages to isolate the subsystem but for a certain period of time until it encounters the influence of the environment.  A common instance of the latter scenario is the presence of shot noise that shows up at random times to disturb the unitary evolution of the subsystem~\cite{Dattagupta:2013}.     
 
 In recent years, there has been an upsurge of interest in investigating time-evolution of a subsystem that is interrupted at random times with non-unitary evolution owing to interactions with either an external environment or a measuring apparatus that has been employed to make measurements on the system. The result would be a series of unitary evolution sequences being stochastically modulated in time,  a scheme in accord with the second scenario painted in the preceding paragraph.  It is pertinent at this stage to point out that experimentally, it is more feasible to generate interactions with an external apparatus at random time intervals, rather than those at regular intervals. Indeed, one would typically use a timer to time the gap between successive interactions, and because the timer would invariably be of finite precision that it would not be possible to ensure that interactions occur at exactly regular time intervals.  In such a scenario, it is reasonable to consider the time intervals to be random variables,  and to repeat the experiment a number of times that would average over these random variables.
        
In the backdrop of the foregoing,  our aim in this paper is to first develop a general formalism employing a stochastic Liouville equation (SLE) for the underlying density operator to study the unitary evolution of a generic quantum system interspersed with non-unitary evolution at random times. The non-unitary evolution is modelled in terms of an interaction superoperator\footnote{In contrast to an ordinary operator that acts on a state vector to yield a state vector,  a superoperator acts on an ordinary operator to give another ordinary operator, see Appendix~\ref{app1}.}.  Our formalism is very general in the sense that it applies to (i) any generic quantum system, (ii) any form of the interaction superoperator,  (iii) any form of the initial density operator, e.g., those characterizing either pure or mixed states,  and (iv) any probability distribution for the random time interval between successive interactions, so long as the intervals are taken to be independent and identically-distributed random variables.  The concerned SLE is known in the literature on spectral line shapes and relaxation, courtesy Clauser and Blume~\cite{Clauser:1971},  but its application in the context of random interruptions of a quantum dynamics has not been attempted hitherto, to the best of our knowledge.   The treatment of Clauser and Blume,  see also Refs. \cite{Dattagupta:1984,Dattagupta:1987}, is adapted here  for deriving a stochastic Liouville equation, as in Refs.~\cite{Kubo:1963} and~\cite{Shushin:2011}, for the Laplace transform of the average density operator at any time during the dynamical evolution.  It is worthwhile to mention that a superoperator formalism similar to the one we invoke is discussed in Ref.~\cite{Shushin:2011}, though its application is restricted to only a two-level system. This is in stark contrast to what we do in our work, namely, to apply the superoperator formalism to discuss a paradigmatic solid-state physics system, the so-called tight-binding model, which involves many sites among which nearest-neighbour hopping of a quantum particle takes place. For this model, we consider two distinct dynamical scenarios, namely, that of projective measurements and stochastic resets, see below;  only the former and not the latter has been considered in Ref.~\cite{Shushin:2011} and that too for a two-level system.  We derive for our case a number of physically relevant results, as we now detail.
       
Having developed our formalism that allows to obtain an analytical closed-form expression for the Laplace transform of the average density operator,  we find it well in order to apply the formalism to a model system in order to derive explicit results demonstrating the physical effect of non-unitary interventions at random times.  To this end,  we choose to study the well-known tight-binding Hamiltonian of solid-state physics that is regarded as a prototypical model for a nano wire~\cite{Dunlap:1986,Dattagupta-resonance}.  The tight-binding system in one dimension,  the so-called tight-binding chain (TBC),  describes the motion of a charged particle on a lattice, wherein the particle is for most times localized on the sites of the lattice, but which because of spontaneous quantum fluctuations makes occasional tunnelling to nearest-neighbour sites.  It is known that the particle at all times has time-dependent probability for it to be found on different sites,  a fact that manifests in the mean-squared displacement (MSD) of the particle about its initial position growing unbounded in time.  

In the framework of the unitary evolution of the TBC interrupted at random times with non-unitary interactions modelled by an interaction superoperator $T$,  we consider in this work  the random intervals between successive interactions to be independent random variables distributed according to an exponential distribution. We consider two representative forms of $T$, one that implements stochastic reset of quantum dynamics~\cite{qreset1}, whereby the density operator is at random times reset to its initial form, and another that implements projective measurements on the system at random times~\cite{Cohen}.  

We employ our developed formalism to derive in the two aforementioned cases an explicit analytical expression for the average occupation probability for the TBC particle to be found on different sites, wherein the average is with respect to the classical randomness associated with the times at which non-unitary interactions mediate the otherwise unitary evolution of the TBC.  We show that in the case of stochastic reset of quantum dynamics,  the system at long times relaxes to a stationary state in which the TBC particle has time-independent probabilities for it to be found on different sites. Concomitantly, the MSD of the particle about its initial location can no longer grow unbounded in time, as is the case with solely unitary evolution, but instead attains a time-independent value.  Thus,  stochastic resets of TBC dynamics lead to localization of the particle at long times,  achieved through classical stochasticity involved with sampling of the random times for implementing resets.  The results obtained in the case of stochastic reset may also be obtained as a simple extension of an alternative theoretical approach developed in Ref.~\cite{qreset1}; we use our approach to obtain the same results with the purpose of benchmarking our method.

 In the case of projective measurements at random times, our exact results for the average site occupation probability of the TBC particle implies that while performing repeated projective measurements at random times to the initial location of the particle,  one may achieve an effective suppression of the temporal decay in the probability of the particle to be found on the initial site.  The amount of suppression is comparable to the one in conventional Zeno effect scenarios~\cite{Misra:1977,Venugopalan:2007},  but which however does not require the experimentally not-so-feasible requirement of performing measurements at exactly regular intervals that is a characteristic of such scenarios. 
        
Given this background, the paper is organised as follows. In Sec. ~\ref{secII}, we develop our formalism via the SLE to treat analytically the unitary evolution of a generic quantum system interspersed with repeated non-unitary interactions at random times.  This is followed up by the application of the formalism to the TBC in Sec.~\ref{secIII}.  Here,  we provide in Sec.~\ref{secIII-a} a brief recollection of some basic properties of the unitary evolution of the TBC,  and then consider the TBC subject to non-unitary interactions at random times; we discuss in turn the case of stochastic resets in Sec.~\ref{secIII-b} and the case of projective measurements in Sec.~\ref{secIII-c}, both for the representative case of exponentially-distributed random time intervals between successive interactions.  The paper ends with conclusions in Sec.~\ref{secIV}.  The two appendices provide respectively a brief reminder on Liouville operator and superoperator and details of numerical implementation of the dynamical protocol of non-unitary interactions at random times for the TBC Hamiltonian.
        
\section{Quantum unitary evolution interspersed with repeated non-unitary interactions at random times: The  average density operator}
\label{secII}

\begin{figure}[ht]
\centering
\includegraphics[scale=1.25]{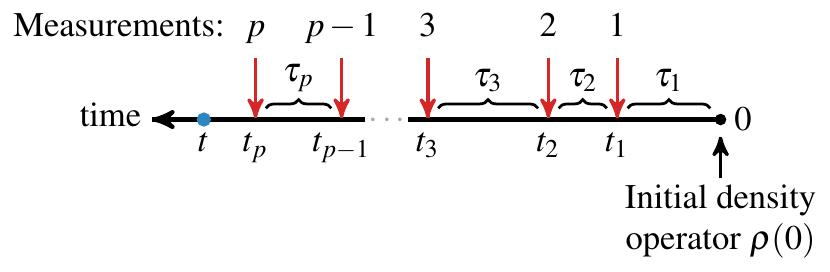}
\caption{The figure shows according to the scheme detailed in Section~\ref{secII} a
typical evolution of a quantum system for time $t$ (shown by a blue dot on the time axis). Starting with a density operator $\rho(0)$,  the evolution involves the following repetitive sequence
of events: unitary evolution for time $\tau_{p'+1} \equiv t_{p'+1} - t_{p'};~p'=0,1,2,\ldots,p-1;~t_0=0$ followed by an instantaneous interaction with the external environment (denoted by red arrows).  The evolution ends with unitary evolution for time duration $t-t_p$.}
\label{fig:evolution-schematic}
\end{figure}

In this section,  we introduce our scheme of dynamical evolution for a general quantum system characterized by a given time-independent Hamiltonian $H$.  We will work in units in which the Planck's constant is unity.  We set up via the stochastic Liouville equation approach a Liouville or a superoperator treatment of the average density operator of the system at an arbitrary time $t>0$ while starting from a given density operator $\rho(0)$ at time $t=0$.   
We consider the quantum system to be undergoing unitary evolution in
time that is interspersed with instantaneous interaction with the external environment occurring at random time instances, up to a certain fixed time. The instantaneous interactions induce non-unitarity in the evolution of the system. 
Specifically, a typical realization of the dynamics for time $t$
involves the following: Starting with $\rho(0)$, a unitary
evolution for a random time is followed by an instantaneous interaction modelled in terms of a given interaction operator $T$.  The result is subject to unitary evolution for another random time,  followed by another instantaneous interaction.  Continuing it this way over the time interval $[\,0,\,t\,]$,  a realization of the  evolution may involve $p \ge 0$ number of instantaneous interactions occurring at
random time instances $t_1,t_2,\ldots,t_p$, with the times $\tau_{p'+1} \equiv t_{p'+1} -t_{p'};~p'=0,1,2,\ldots,p-1;~\,t_0=0$ between
successive interactions being random variables that we choose to  be sampled independently from a
common distribution $p(\tau)$.  The evolution ends with unitary evolution for time duration $t-t_p$.
In Fig.~\ref{fig:evolution-schematic}, we show a schematic diagram of the aforementioned dynamical protocol. 

The average density
operator (averaged over different realizations of the dynamics detailed above) at time $t$ reads
\begin{align}
\overline{\rho}(t) &= \sum_{p=0}^\infty \int_0^t {\rm d}t_p \int_0^{t_p} {\rm d}t_{p-1}\ldots \int_0^{t_3}{\rm d}t_2 \int_0^{t_2} {\rm d}t_1 \, F(t-t_p) \, \ee^{-\im\mathcal{L}(t-t_p)} \nonumber \\
&\hskip30pt \times Tp(t_p-t_{p-1})\ee^{-\im\mathcal{L}(t_p-t_{p-1})}T\ldots
Tp(t_2-t_1)\ee^{-\im\mathcal{L}(t_2-t_1)}T p(t_1)e^{-\im\mathcal{L}t_1}\rho(0) 
\label{eq:rho-evolution-0} \\
&= U(t)\rho(0),
\label{eq:rho-evolution-1}
\end{align}
where $U(t)$ is a superoperator in the sense that it acts on one operator, namely, $\rho(0)$, to yield another operator given by the left hand side.  Appendix~\ref{app1} provides a reminder on superoperators. 
In Eq.~(\ref{eq:rho-evolution-0}), the quantity $F(t)$ is the probability
for no interaction to occur during time $t$. Equation~(\ref{eq:rho-evolution-0}) is easy to interpret (refer Fig.~\ref{fig:evolution-schematic}).  Starting at time $t_0=0$ with the density operator $\rho(0)$,  a realization of the dynamics involving the time sequence $\{t_0=0,t_1,t_2,\ldots,t_p\}$ and observation at time $t>t_p$ would comprise unitary evolution following the evolution equation
\begin{align}
\rho(t'>t'')=\ee^{-\im \mathcal{L}(t'-t'')}\rho(t'')=\ee^{-\im H(t'-t'')}\rho(t'')\ee^{\im H(t'-t'')};
\label{eq:rho-evolution}
\end{align}
the unitary evolution occurs for time $\tau_1=t_1-t_0$,  the probability of which is $p(\tau_1)$,  and this is followed by an instantaneous interaction with the external environment,  again unitary evolution for $\tau_2=t_2-t_1$, followed by an instantaneous interaction, and so on, until the last interaction in the sequence performed at time $t_p$, and then unitary evolution until the time $t$ of observation, that is, for duration $t-t_p$.  No interaction should happen during this duration $t-t_p$, the probability of which is provided by the factor $F(t)$.  Integrating over all possible values of $t_1,t_2,\ldots,t_p$ and summing over all possible values of $p$ yield the average density operator in Eq.~(\ref{eq:rho-evolution-0}).

In order to proceed,  let us now recall the result for Laplace transform of a series of
convolutions. The convolution between two functions $g_1(t)$ and
$g_2(t)$ over an interval $[0,t]$ is
defined as $g_1*g_2\equiv\int_0^t {\rm d}\tau~g_1(\tau)g_2(t-\tau)$. We then
have $\mathfrak{L}(g_1*g_2)=\mathfrak{L}(g_1)\mathfrak{L}(g_2), $
with $\mathfrak{L}$ denoting the Laplace transform operator. Generalizing the result to any number of convolutions, and applying it to
Eq.~(\ref{eq:rho-evolution-0}),  we get
\begin{align}
\widetilde{\overline{\rho}}(s) &\equiv \mathfrak{L}(\overline{\rho}(t)) = \widetilde{U}(s)\rho(0); \label{eq:rho-in-lap} \\
\widetilde{U}(s) &= \mathfrak{L}\big(F(t)\ee^{-\im \mathcal{L}t}\big)\sum_{p=0}^\infty \Big[T\mathfrak{L}\big(p(t)\ee^{-\im \mathcal{L}t}\big)\Big]^{p}=\frac{\mathfrak{L}(F(t)\ee^{-\im \mathcal{L}t})}{\mathbb{I}-T\mathfrak{L}(p(t)\ee^{-\im \mathcal{L}t})},
\label{eq:Ut-LT}
\end{align}
where in the last step we have used the operator identity
\begin{align}
(A - B)^{-1} = A^{-1} + A^{-1}B(A-B)^{-1}  \label{eq:mat-identity-inv}
\end{align}
with $A = \mathbb{I}$ and $B = T\mathfrak{L}(p(t)\exp(-\im \mathcal{L}t))$.
Equation~(\ref{eq:rho-in-lap}) is a straightforward extension of the formalism developed by Clauser-Blume~\cite{Clauser:1971} to a general $p(t)$. 

A few relevant comments are in order regarding the result~(\ref{eq:rho-in-lap}).  Firstly,  the result is exact and allows on performing inverse Laplace transform to obtain the average density operator $\overline{\rho}(t)$.  Secondly,  the result is very general and applies to (i) any generic quantum system described by a time-independent Hamiltonian, (ii) any form of the interaction operator $T$,  (iii) any form of the density operator, e.g., those characterizing either pure or mixed states,  and (iii) any probability distribution $p(t)$.  

We will now show two explicit applications of the developed formalism, by specializing to the case of the tight-binding chain and to an exponential $p(\tau)$ given by 
\begin{align}
p(\tau)=\lambda\ee^{-\lambda\tau},
\label{eq:ptau-exponential}
\end{align}
where $\lambda>0$ is the inverse of the average time between two
successive interactions.   In this case, we have 
\begin{align}
F(t)=\int_t^\infty {\rm d}\tau~p(\tau)=\lambda \int_t^\infty {\rm
d}\tau~\exp(-\lambda \tau)=\exp(-\lambda t). 
\end{align}
Then, using the result
\begin{align}
\mathfrak{L}(t^{n} \ee^{-at}u(t)) = \int_{0}^{\infty} {\rm d}t~\ee^{-(s+a)t}t^n =\frac{n!}{(s+a)^{n+1}}, \label{eq:LT-1}
\end{align}
where $n \in [0,1,\ldots,\infty)$ and $u(t)$ is the unit-step function, one may obtain from Eq.~(\ref{eq:Ut-LT}) that
\begin{align}
 \widetilde{U}(s)=\left[(s+\lambda)\mathbb{I}+\im\mathcal{L}-T\lambda\right]^{-1}.
        \label{eq:Us}
\end{align}
Using the operator identity~\eqref{eq:mat-identity-inv}, we may expand $\widetilde{U}(s)$ in Eq.~(\ref{eq:Us}) in an infinite series as
\begin{align}
\widetilde{U}(s) =\widetilde{U}_0(s)+\lambda
\widetilde{U}_0(s)T\widetilde{U}_0(s)+\lambda^2 \widetilde{U}_0(s) T
\widetilde{U}_0(s) T \widetilde{U}_0(s)+\ldots,
\label{eq:U-expansion}
\end{align}
where we have
\begin{align}
\widetilde{U}_0(s) \equiv [(s+\lambda)\mathbb{I}+\im\mathcal{L}]^{-1} .\label{eq:U0s} 
\end{align}

\section{Application to the tight-binding chain (TBC)}
\label{secIII}

\subsection{The model}
\label{secIII-a}

The tight-binding model (TBM)~\cite{Dunlap:1986,Dattagupta-resonance} is a textbook description of a quantum solid in which an electron is assumed mostly localized on the sites of a lattice but because of spontaneous quantum fluctuations makes occasional tunnelling to nearest-neighbour sites.  Specifically, the tight-binding chain (TBC),  referring to the TBM in one dimension, is characterized by a Hamiltonian modelling the motion of a quantum particle, e.g., an electron, between the nearest-neighbour sites of a one-dimensional lattice. The Hamiltonian reads
\begin{align}
H  = - \frac{\Delta}{2} \sum_{n=-\infty}^\infty \Big( |n\rangle  \langle n+1| + |n+1\rangle \langle n| \Big),
\label{eq:TBC-Hamiltonian}
\end{align}
where $\Delta > 0$ is the nearest-neighbour intersite hopping integral,  and $|n\rangle$ is the so-called Wannier state denoting the state of the particle when located on site $n$.  The Wannier states form a set of complete basis states satisfying $\langle m|n\rangle=\delta_{mn}$ and $\sum_{m=-\infty}^\infty |m\rangle \langle m|=\mathbb{I}$, the identity operator. Here,  we have taken the lattice spacing to be unity, without loss of generality.  

The TBM plays a critical role in the band theory of solids. With the advent of mesoscopic physics,  the TBM finds relevance in discussing one-dimensional nanowires~\cite{Hirsch:1999, Zhang:2000,Sinova:2015}.  In this subsection, we collect relevant results for the quantum mechanics of a tight-binding chain. Although most of these results are already documented in the literature vide the work of Dunlap and Kenkre~\cite{Dunlap:1986}, we nevertheless systematically collate them here in order to apply the formalism developed in Sec.~\ref{secII} later in the paper.
        
A particle moving under the Hamiltonian~(\ref{eq:TBC-Hamiltonian}) has a motion akin to the ballistic motion of a classical particle. To see this explicitly, we introduce the operators
\begin{align}
K \equiv \sum_{n=-\infty}^\infty |n\rangle  \langle n+1|, ~~K^\dagger \equiv \sum_{n=-\infty}^{\infty} |n+1\rangle \langle n|,   
\end{align}
in terms of which we have      
\begin{align}   
H =-\frac{\Delta}{2}  (K + K^\dagger). \label{eq:model-1}
\end{align}                                                                  
Note that $K$ and $K^\dagger$ commute with each other,  with their product being $KK^\dagger=K^\dagger K=\mathbb{I}$.  Also, one has $K^2=\sum_{n=-\infty}^\infty |n\rangle\langle n+2|$ and $(K^\dagger)^2=\sum_{n=-\infty}^\infty |n+2\rangle\langle n|$.  That $K$ and $K^\dagger$ commute implies that we can find a representation in which $K$ and $K^\dagger$ can be simultaneously diagonalized. Such a representation is elicited through the so-called Bloch states, defined in the reciprocal momentum-space as
\begin{align}
|k\rangle \equiv \frac{1}{\sqrt{2\pi}}\sum_{n=-\infty}^\infty \ee^{-{\rm i}nk}|n\rangle,
\end{align}
so that we have $|n\rangle = (1/\sqrt{2\pi})\int_{-\pi}^\pi {\rm d}k\, \exp({\rm i}nk)|k\rangle$.  It is easily verified that 
\begin{align}
K |k\rangle = \ee^{-{\rm i}k} |k\rangle, ~~K^\dagger |k\rangle = \ee^{{\rm i}k} |k\rangle,
\label{eq:Bloch}
\end{align}             
thus proving that the Bloch state $|k\rangle$ is indeed an eigenstate of $K$ and $K^\dagger$, with eigenvalues $\exp (- {\rm i} k)$ and $\exp ({\rm i} k)$, respectively.  We therefore have
\begin{align}
\ee^{\im H t}|m'\ra&=\frac{1}{\sqrt{2\pi}}\int_{-\pi}^\pi {\rm d}k~\ee^{\im
km'}\ee^{\im Ht}|k\ra=\frac{1}{\sqrt{2\pi}}\int_{-\pi}^\pi {\rm d}k~\ee^{\im
km'}\ee^{-\im \Delta t\cos k}|k\ra,
\label{eq:Heqn-1}
\end{align}
where in obtaining the last equality, one uses Eq.~(\ref{eq:Bloch}).
We therefore have on using $\langle k'|k\rangle=\delta(k-k')$ that
\begin{align}
\la m|\ee^{\im H t}|m'\ra=\frac{1}{2\pi}\int_{-\pi}^\pi {\rm
d}k~\ee^{-\im k (m-m') - \im \Delta t \cos k}.
\label{eq:Heqn}
\end{align}

Let us now ask the question: what is the probability $P_m(t)$ that the particle is on site $m$ at time $t>0$, given that the particle was on site $n_0$ at time $t=0$? By definition, we have
\begin{align}
P_m(t)=\langle m|\rho(t)|m\rangle. \label{eq:Pmt-as-rho-element}
\end{align}
Here,  $\rho(t)$ is the density operator of the system, whose time evolution follows the usual dynamics 
\begin{align}
\rho(t)=\ee^{-\im \mathcal{L}t}\rho(0)=\ee^{-\im Ht}\rho(0)\ee^{\im Ht}.
\label{eq:rho-evolution-again}
\end{align}
Equation~(\ref{eq:rho-evolution-again}) implies unitary evolution of the density operator that preserves its trace: ${\rm Tr}[\rho(t)]=1~\forall~t$.   
In our case,  we have $\rho(0)=|n_0\rangle \langle n_0|$, so that it may be shown on using Eq.~(\ref{eq:Heqn}) that
\begin{equation}
P_m(t)=\frac{1}{(2\pi)^2} \int_{-\pi}^\pi {\rm d}k \int_{-\pi}^\pi {\rm d}k' ~\ee^{{\rm i} (m-n_0) (k-k')} ~\ee^{-{\rm i}\Gamma_{kk'}t},
\label{eq:Pmt-bare}
\end{equation}
with
\begin{align}
\Gamma_{kk'}\equiv \Delta (\cos k'-\cos k).
\label{eq:Gammakk'}
\end{align}
It may be checked that as desired,  $\sum_{m=-\infty}^\infty P_m(t)=1$.  Using the identities
$\exp(-\im \Delta t \cos k) = \sum_n \exp(-\im n\pi /2 + \im nk) J_n(\Delta t)$~\cite{Dunlap:1986}, with $J_n(x)$ the Bessel function of the first kind,  and $\int_{-\pi}^\pi {\rm d}k~\exp(\im(m-n)k)=2\pi \delta_{mn}$, 
one may straightforwardly show that
\begin{align}
\frac{1}{(2\pi)^2}\int_{-\pi}^\pi {\rm
d}k\int_{-\pi}^\pi {\rm d}k'~\ee^{\im m(k-k')} \ee^{-\im
\Gamma_{kk'}t}=J_m^2(\Delta t).
\label{eq:P0-3}
\end{align}
Then, from Eq.~\eqref{eq:Pmt-bare}, we get~\cite{Note:Bessel}
\begin{align}
P_m(t)=J_{m-n_0}^2(\Delta t),
\label{eq:Pmt-bare-Bessel}
\end{align}
implying that $\Delta$ sets the intrinsic time scale in the model.  Using the result that in the limit $\nu \to \infty$ we have for $x \ne 0$ and fixed that $J_\nu(x) \sim (1/\sqrt{2\pi \nu})(\ee \, x/(2\nu))^\nu$~\cite[Eq.~10.19.1]{DLMF-NIST},  we get 
\begin{align}
P_m(t) \sim \left(\frac{\ee \, \Delta~t}{2 |m-n_0|}\right)^{2|m-n_0|}; \quad |m-n_0| \to \infty.
\label{eq:Pmt-bare-Bessel-limit}
\end{align}

The discrete Fourier transform of the function $P_m(t)$, given by $\widetilde{P}(k,t) \equiv 1/(\sqrt{2\pi})\sum_{m=-\infty}^\infty \exp(-{\rm i}mk)P_m(t)$, is obtained as
\begin{align}
\widetilde{P}(k,t)&=\frac{1}{(2\pi)^{3/2}}\int_{-\pi}^\pi {\rm d}k' ~ \ee^{-{\rm i}n_0 k} ~\ee^{{\rm i}\Delta t \left(\cos (k+k')- \cos k'\right)},
\label{eq:Pmt-Fourier}
\end{align} 
using $\sum_{m=-\infty}^\infty \exp(-{\rm i}(k+k_1-k_2)m)=2\pi \delta(k+k_1-k_2)$.

Equation~(\ref{eq:Pmt-bare}) implies that the particle while starting from the site $n_0$ eventually spreads out to other sites of the lattice.  The average displacement from the initial location $n_0$ is given by 
\begin{align}
\mu \equiv \sum_{m=-\infty}^\infty (m-n_0)P_m(t)=0, \label{eq:MEAN-lam0}
\end{align}
since Eq.~(\ref{eq:Pmt-bare}) implies that $P_m(t)$ is invariant under $m-n_0 \to -(m-n_0)$. Let us then compute the mean-squared displacement (MSD) of the particle about $n_0$:
\begin{align}
S(t)\equiv \sum_{m=-\infty}^\infty (m-n_0)^2 P_m(t)=\sum_{m=-\infty}^\infty m^2 P_m(t)+n_0^2-2n_0 \sum_{m=-\infty}^\infty m P_m(t), 
\label{eq:MSD-bare}
\end{align}
where we have used $\sum_{m=-\infty}^\infty P_m(t)=1$.  Now,  we have $
\sum_{m=-\infty}^\infty m P_m(t)=\sum_{m=-\infty}^\infty (m-n_0+n_0)P_m(t)=n_0$,  so that $S(t)=\sum_{m=-\infty}^\infty m^2 P_m(t)-n_0^2$.  In terms of $\widetilde{P}(k,t)$, we get $
\sum_{m=-\infty}^\infty m^2 P_m(t)=-\sqrt{2\pi}{\rm d}^2 \widetilde{P}(k,t)/{\rm d}k^2|_{k=0}$, using which it may be shown straightforwardly that
\begin{align}
S(t)=\frac{\Delta^2 t^2}{2}. \label{eq:MSD-lam0}
\end{align}
Thus, we see that the MSD has a quadratic time dependence similar to the case of the ballistic motion of a free classical particle but with a `quantal' speed equal to $\Delta /\sqrt{2}$.  In interpreting Eq.~(\ref{eq:MSD-lam0}),  note that we have set the lattice spacing to unity at the outset.

\subsection{The model subject to stochastic resets at random times}
\label{secIII-b}

Here, we will consider the form of $T$ that implements stochastic resets of the TBC system at random times.  We formulate our problem thus: The quantum particle is taken to begin its journey at time $t = 0$ from an arbitrary site $n_0$ of the TBC, so that the corresponding density operator is $\rho(0)=|n_0\rangle \langle n_0|$. It then streams unitarily under the Hamiltonian (\ref{eq:TBC-Hamiltonian}) until a random time $t_1$ at which an instantaneous reset takes place of the density operator onto the form $|{\cal N}\ra \la {\cal N}|$ corresponding to a \textit{complete} collapse of the state vector of the system onto the state $|{\cal N}\ra$ corresponding to an arbitrary site ${\cal N}$.   Subsequent to the first reset,  the sojourn of the particle begins anew from the site ${\cal N}$ and an unitary evolution follows until time $t_2\, (t_2 > t_1)$, again random, when a second reset takes place and subsequent to which the particle evolution is renewed a second time.This ongoing `renewal' process, much akin to the random walk scenario of Montroll \textit{et al.}~\cite{Montroll:1965, Scher:1975,Weiss:1994}, continues at a sequence of random times,  and finally,  we may ask at time $t$ about the probability $P_m(t)$ for the particle to be on an arbitrary site $m$ of the lattice.  We note that for exponential $p(t)$,  we may view the dynamics in a manner similar to the one discussed in Ref.~\cite{qreset1}: in a small time interval ${\rm d}t$,  the wave vector of the system becomes either the state $|{\cal N}\ra$ with probability $\lambda{\rm d}t$ (i.e., there is a reset to the state $|{\cal N}\ra$), or, evolves unitarily under the Hamiltonian $H$ with probability $1-\lambda{\rm d}t$~\cite{note-last-reset}; consequently, reset to the state $|{\cal N}\ra$ is  indeed a probabilistic event in time.  

Stochastic reset of quantum dynamics, involving quantum dynamics interrupted at random times with reset to a given state,  has been a topic of much current interest~\cite{qreset1}, following recent extensive research on resetting in the classical domain.  This emerging field of research was initialized in Ref.  \cite{Evans:2011} in the context of classical diffusion of a single particle that is reset to its initial location at random times.  A remarkable revelation of the study is the following.  In the absence of resetting, the distribution of the particle location is an ever-spreading Gaussian centered at the initial location of the particle and so does not have a stationary state.  By contrast,  introducing resetting into the dynamics leads to a time-independent probability distribution at long times and the emergence of a nonequilibrium stationary state.  Reference \cite{Evans:2011} is considered a landmark contribution in recent times, which has really ushered in a new beginning in studies of stochastic processes.  Starting with the example of a single diffusing particle, subsequent focus has been on a wide variety of dynamical scenarios of both theoretical and practical relevance.  To name a few,  we may cite in the context of a single particle the example of its diffusive motion in a bounded domain~\cite{reset1} and in a potential~\cite{reset2}, and the examples of continuous-time random walk~\cite{reset3,reset4}, L\'{e}vy~\cite{reset5} and exponential constant-speed flights~\cite{reset6}.  Resetting has also been studied in interacting particle systems such as fluctuating interfaces~\cite{reset7,reset8} and reaction-diffusion models~\cite{reset9}.  Stochastic resets of quantum dynamics studied in Ref.~\cite{qreset1} considered resetting of several representative closed quantum systems at constant rate $r$,  to show that any generic observable never thermalizes under such dynamics for any $r$.  We refer the reader to the recent review \cite{Evans:2020} on the topic of resetting.

We now set out to obtain the probability $\overline{P}_m(t)$ for the particle to be on site $m$ at time $t$ while evolving according to our scheme of repeated resets at random times.  The overline denotes as in Eq.~(\ref{eq:rho-evolution-0}) an average over different realizations of the dynamical protocol.  To this end,  we will employ the formalism developed in Sec.~\ref{secII}. 
 In our case,  we have
\begin{align}
\rho(0)=|n_0\rangle \langle n_0|,
\label{eq:rho0-TBC}
\end{align}
while the projection superoperator $T$ projects the instantaneous density operator onto the form $|{\cal N}\rangle \langle {\cal N}|$.
The matrix elements of the superoperator $\widetilde{U}_0(s)$ defined in Eq.~\eqref{eq:U0s} may be obtained by referring to Appendix~\ref{app1}, and by using Eqs.~\eqref{eq:LT-1} and~\eqref{eq:Heqn} as
\begin{align}
(  m n |  \widetilde{U}_0(s) | m'n' )&= ( mn|  \int_0^\infty {\rm d}t~  \ee^{-(s+\lambda){\mathbb{I}}t - \im \mathcal{L} t} | m'n' ) 
=\int_0^\infty {\rm d}t~\ee^{-(s+\lambda)\mathbb{I}t}\la m|\ee^{-\im Ht}|m'\ra \la n'|\ee^{\im Ht}|n\ra  \nonumber  \\
&= \frac{1}{(2\pi)^2} \int_{-\pi}^\pi {\rm
d}k\int_{-\pi}^\pi {\rm d}k'~\ee^{\im (m-m')k-\im (n-n')k'} \frac{1}{s+\lambda+\im
\Gamma_{kk'}}.\label{eq:U-identity-0}
\end{align}
It then follows that
\begin{align}
\sum_m (mm|\widetilde{U}_0(s)|nn)=\frac{1}{s+\lambda},
\label{eq:U-identity-3}
\end{align}
which may be shown by using $\sum_{m = -\infty}^{\infty} \exp(\im m (k-k'))=2\pi \delta(k-k')$.

We want the form of $T$ to be such that subsequent to a reset at any time instant $t$,  the density operator becomes
$\rho_+(t)=T\rho_-(t)=\ket{{\cal N}}\bra{{\cal N}}$, with the trace of the operators satisfying ${\rm Tr}[\rho_+(t)]={\rm Tr}[\rho_-(t)]=1~\forall~t$ and $\rho_-(t)$ denoting the
density operator prior to the reset. This is
guaranteed with the form 
\begin{align}
(n_1 n_1'|T|n_2 n_2')=\delta_{n_1 n_1'}\delta_{n_2n_2'}\delta_{n_1 {\cal N}}.
\label{eq:Tmatrix}
\end{align}
Indeed, then, we have
\begin{align} 
\la n|T\rho_-(t)|n' \ra &=\sum_{n_3,n_4}(nn'|T|n_3n_4)\bra{n_3}\rho_-(t)\ket{n_4}=\delta_{nn'}\delta_{n{\cal N}}\sum_{n_3}\bra{n_3}\rho_-(t)\ket{n_3}=\delta_{nn'}\delta_{n{\cal N}},
\end{align}
implying that $T\rho_-(t)=\ket{\cal N}\bra{\cal N}$; here we have used the fact
that ${\rm Tr}[\rho_-(t)]=1~\forall~t$.

Now,  by definition, we have
\begin{align}
\overline{P}_m(t)=\langle m|\overline{\rho}(t)|m\rangle.
\label{eq:Pmt-equation}
\end{align}
Denoting $\overline{P}_m(t)$ in the Laplace domain by $\widetilde{\overline{P}}_m(s)$, one obtains from Eqs.~\eqref{eq:rho-in-lap} and~\eqref{eq:U-expansion} that
\begin{align}
 \widetilde{\overline{P}}_m(s)  &= \la m | \widetilde{U}_0(s)\rho(0) +\lambda 
\widetilde{U}_0(s)T\widetilde{U}_0(s) \rho(0)+\lambda^2 \widetilde{U}_0(s) T
\widetilde{U}_0(s) T \widetilde{U}_0(s) \rho(0)+\ldots |m\ra \nonumber \\ 
& \equiv \sum_{p = 0}^{\infty} \widetilde{\overline{P}}_m^{(p)}(s) ,
\label{eq:Pms}
\end{align}
where $\widetilde{\overline{P}}_m^{(p)}(s);~p \in [0,\infty)$ is the $p$-th term of the infinite series,  which corresponds to $p$ instantaneous interactions.

From Eq.~(\ref{eq:Pms}), we get
\begin{align}
\widetilde{\overline{P}}_m^{(0)}(s)&=\la m|\widetilde{U}_0(s)\rho(0)| m \ra 
=\sum_{n,n'}(mm|\widetilde{U}_0(s)|nn')\bra{n}\rho(0)\ket{n'} 
= (mm|\widetilde{U}_0(s)|n_0n_0) \nonumber \\
&=\frac{1}{(2\pi)^2}\int_{-\pi}^\pi {\rm
d}k\int_{-\pi}^\pi {\rm d}k'~\ee^{\im (m-n_0)(k-k')} \frac{1}{s+\lambda+\im
\Gamma_{kk'}}, \label{eq:P0s-1}
\end{align}
where we have used Eqs.~(\ref{eq:rho0-TBC}) and~\eqref{eq:U-identity-0}.
The quantity $\overline{P}_m^{(0)}(t)$ may be read off from Eq.~(\ref{eq:P0s-1}) to be
\begin{align}
\overline{P}_m^{(0)}(t) &=\frac{\ee^{-\lambda t}}{(2\pi)^2}\int_{-\pi}^\pi {\rm
d}k\int_{-\pi}^\pi {\rm d}k'~\ee^{\im (m-n_0)(k-k')} \ee^{-\im \Gamma_{kk'}t}= \ee^{-\lambda t}J_{m-n_0}^2(\Delta t),
\label{eq:P0-2}
\end{align}
where we have used Eq.~(\ref{eq:P0-3}).  From Eqs.~(\ref{eq:P0s-1}) and~(\ref{eq:P0-2}), we get
\begin{align}
\mathfrak{L}((mm|\widetilde{U}_0(s)|n_0n_0))=\ee^{-\lambda t}J_{m-n_0}^2(\Delta t).
\label{eq:U0s-result}
\end{align}

Similarly, from Eq.~\eqref{eq:Pms}, we get
\begin{align}
\widetilde{\overline{P}}_m^{(1)}(s)
&=\lambda\sum_{\substack{n_1,n_2,n_3,\\ n_4,n_5,n_6}}
(mm|\widetilde{U}_0(s)|n_1n_2)(n_1n_2|T|n_3n_4)   (n_3n_4|\widetilde{U}_0(s)|n_5n_6)\la
n_5|\rho(0)|n_6\ra \nonumber \\[1ex]
&=\lambda\sum_{n_3} (mm|\widetilde{U}_0(s)|{\cal N}{\cal N})(n_3n_3|\widetilde{U}_0(s)|n_0n_0) \nonumber \\
&=\frac{\lambda}{(2\pi)^2}\int_{-\pi}^\pi {\rm
d}k_1\int_{-\pi}^\pi {\rm d}k_2~\ee^{\im (m-{\cal N})(k_1-k_2)}\frac{1}{(s+\lambda)(s+\lambda+\im \Gamma_{k_1k_2})} ,
\end{align}
where in obtaining the last step, we have used
Eqs.~\eqref{eq:U-identity-0} and~(\ref{eq:U-identity-3}).  Using the fact that the Laplace transform of a convolution of two functions is given by the product of their individual Laplace transforms,  we may write 
\begin{align}
\widetilde{\overline{P}}_m^{(1)}(s)&=\lambda \mathfrak{L}\Big(\int_0^t {\rm d}t'\,\ee^{-\lambda (t-t')} \frac{1}{(2\pi)^2}\int_{-\pi}^\pi {\rm
d}k_1\int_{-\pi}^\pi {\rm d}k_2~\ee^{\im (m-{\cal N})(k_1-k_2)} \ee^{-(\lambda+\im \Gamma_{k_1k_2})t'}\Big)\nonumber \\
&=\lambda \mathfrak{L}\Big(\ee^{-\lambda t} \int_0^t {\rm d}t'\,J_{m-{\cal N}}^2(\Delta t')\Big),
\end{align}
where we have used Eq.~(\ref{eq:P0-3}). We thus arrive at the result
\begin{align}
\overline{P}_m^{(1)}(t)=\lambda \ee^{-\lambda t}\int_0^t {\rm d}t'~J^2_{m-{\cal N}}(\Delta t').
\end{align}

Proceeding in a similar manner, we get
\begin{align}
\widetilde{\overline{P}}_m^{(2)}(s) &=\frac{\lambda^2}{(2\pi)^2}\int_{-\pi}^\pi {\rm
d}k_1\int_{-\pi}^\pi {\rm d}k_2~\ee^{\im (m-{\cal N})(k_1-k_2)} \frac{1}{(s+\lambda)^2(s+\lambda+\im
\Gamma_{k_1k_2})}.
\end{align}
 Using the result on Laplace transform of a convolution and Eq.~(\ref{eq:P0-3}),  the last equation may be written as
\begin{align}
\widetilde{\overline{P}}_m^{(2)}(s)=\lambda^2 \mathfrak{L}\Big(\ee^{-\lambda t} \int_0^t {\rm d}t'\,(t-t')~J_{m-{\cal N}}^2(\Delta t')\Big)
\end{align}
yielding 
\begin{align}
\overline{P}_m^{(2)}(t)=\lambda^2~\ee^{-\lambda t}\int_0^t {\rm d}t'~(t-t')J^2_{m-{\cal N}}(\Delta t').
\end{align}

\begin{figure*}
\centering
\includegraphics[scale=1.15]{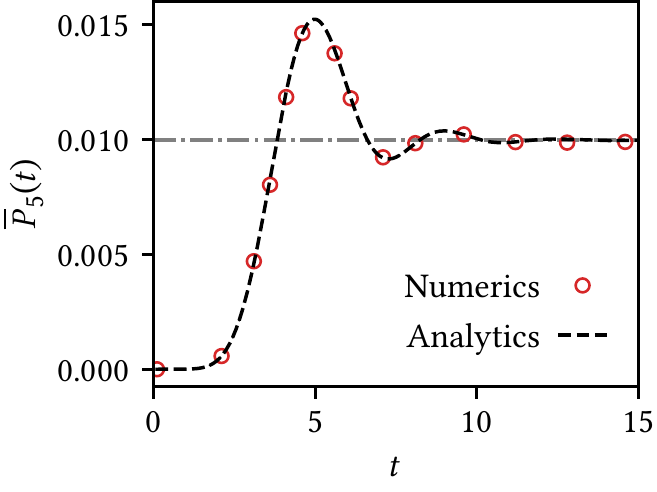} \hskip5pt
\includegraphics[scale=1.15]{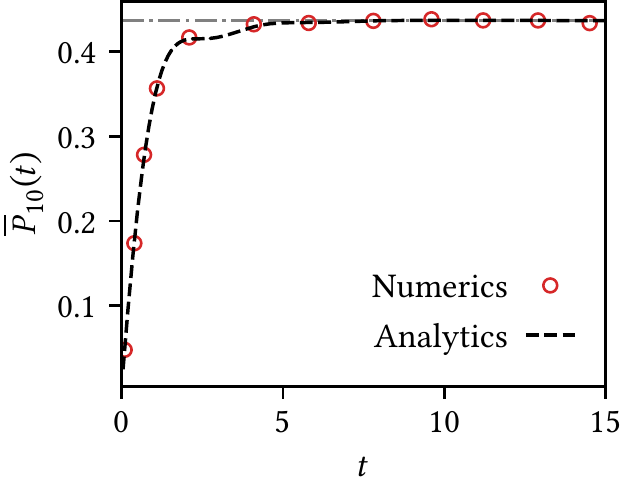}
        \caption{The figures show for the tight-binding model~\eqref{eq:model-1} subject to repeated resets at random times the dependence of 
        the average site occupation probability 
        $\overline{P}_m(t)$ on $t$ for two values of $m$, namely, $m=5$ (left panel) and $10$ (right panel). The initial location of the particle is $n_0 =1$, while the location of the detector is ${\cal N}=10$; also,  we have $ \Delta = 1$, and 
        $\lambda=0.5$.   In the plots, the dashed line in black depicts the analytical result given in Eq.~\eqref{eq:result-rho}, while the points are obtained from numerical implementation of the dynamics and involving averaging over $4\times 10^3$ realizations and a periodic lattice of $N=30$ sites; see Appendix \ref{app2} for details on numerical implementation.  The dash-dotted line in grey shows the long-time value of $\overline{P}_m(t)$ given in Eq.~\eqref{eq:Pst-long-time}.  The plots show a very good agreement between the theoretical and the numerical result. }
        \label{fig:meas}
\end{figure*}

The general expression may thus be written as
\begin{align}
\overline{P}_m^{(p)}(t)=\lambda^p ~\ee^{-\lambda t} \int_0^t \!\!{\rm
d}t'~\frac{(t-t')^{p-1}}{(p-1)!}~J^2_{m-{\cal N}}(\Delta
t'); \quad p\in[1,\infty),
\end{align}
so that inverting Eq.~\eqref{eq:Pms} back to the time domain and using the above equation, one gets~\cite{Note:Bessel}
\begin{align}
        \overline{P}_m(t) =\ee^{-\lambda t}J_{m-n_0}^2(\Delta t)+ \lambda \int_0^t  {\rm
d}t'~\ee^{-\lambda t'}J_{m-{\cal N}}^2(\Delta t').
\label{eq:result-rho}
\end{align}
Equation~(\ref{eq:result-rho}) is a key result of the paper, yielding the average probability at time $t$ to be on site $m$ for a quantum particle starting at site $n_0$ and evolving under Hamiltonian (\ref{eq:TBC-Hamiltonian}) subject to repeated resets at random times. This result may be contrasted with Eq.~(\ref{eq:Pmt-bare-Bessel}) that holds in the absence of resets.
Summing over $m$ on both sides of Eq.~(\ref{eq:result-rho}),  and using the results $J_{-m}(x)=(-1)^m J_m(x)$ and $1=J^2_0(x)+2\sum_{m=1}^\infty
J^2_m(x)$, we get $\sum_{m=-\infty}^\infty \overline{P}_m(t)= \exp(-\lambda t)+ \lambda \int_0^t  {\rm d}t' \exp(-\lambda t')=1$, as desired. 

The result~(\ref{eq:result-rho}),  which may be obtained as a simple extension of formula (3) in Ref.~\cite{reset7}, has a simple physical interpretation.  The first term on the right hand side arises from those realizations of evolution for time $t$ that did not involve a single reset (the probability for which is $\exp(-\lambda t)$) \cite{note-last-reset},  so that the corresponding contribution to $\overline{P}_m(t)$ would be given by the product of the probability of no reset for time $t$ with the probability to be on site $m$ at time $t$ in the absence of any reset and while starting from site $n_0$ at time $t=0$.  The latter quantity is obtained from our analysis in~\ref{secII} as equal to $J_{m-n_0}^2(\Delta t)$, see Eq.~(\ref{eq:Pmt-bare-Bessel}). In order to understand the second term on the right hand side, let us appreciate that every time there is a reset,  the system collapses to the state $|{\cal N}\rangle$ and its evolution starts afresh from site ${\cal N}$.  Consequently, at time $t$, what should matter is when has been the last reset,  and the corresponding contribution to $\overline{P}_m(t)$ would be given by the product of the probability at time $t$ 
of the last reset to occur in the interval $[\, t',\, t'-{\rm d}t'\,]$, with $t'\in [\, 0, \, t \,]$, 
with the probability in the absence of any reset for the particle to be on site $m$ owing to evolution for time duration $t-t'$ while starting from site ${\cal N}$.  The former probability is given by $\lambda {\rm d}t'\, \exp(-\lambda(t-t'))$ \cite{note-last-reset}.  We thus get the contribution as equal to $\lambda \int_0^t {\rm d}t'\, \exp(-\lambda(t-t'))~J_{m-{\cal N}}^2(\Delta(t-t'))$, to which effecting a change of integration variable directly yields the second term on the right hand side of Eq.~(\ref{eq:result-rho}).  An interpretation similar to the foregoing leads in the case of resetting in the classical domain~\cite{Evans:2020} to a result for the probability distribution similar in form to the one in Eq.  (\ref{eq:result-rho}). We note in passing that the result~(\ref{eq:result-rho}) may also be obtained by using the formalism of Ref.~\cite{qreset1}.  However, our motivation here is not to obtain only the reset result~(\ref{eq:result-rho}), but to put forward a general formalism as in Sec.~\ref{secII} to treat unitary evolution interspersed with non-unitary interactions at random times modelled by \textit{any} interaction superoperator $T$, and the reset case is but a special case of this scenario.

Now, using the result $\sum_{m=-\infty}^{\infty} m \,J^2_{m-n_0}(x) = n_0$ (see the paragraph following Eq.~(\ref{eq:MSD-bare})), one obtains from Eq.~\eqref{eq:result-rho} the average displacement of the particle from the initial position $n_0$ as
\begin{align} 
\overline{\mu}(t) \equiv \sum_{m=-\infty}^{\infty} (m-n_0) ~\overline{P}_{m}(t) = ({\cal N} - n_0) \Big[ 1-\ee^{-\lambda t } \Big] , \label{eq:MEAN-lam-non0}
\end{align}
which in the limit $\lambda \to 0$ reproduces as expected the result~\eqref{eq:MEAN-lam0}.
Next,  by using the result $\sum_{m=-\infty}^{\infty} m^2 \,J^2_{m}(x) = x^2/2$ (see Eqs.~(\ref{eq:MSD-bare}) and~(\ref{eq:MSD-lam0})), the MSD of the particle about the initial location $n_0$ is obtained from Eq.~\eqref{eq:result-rho} as
\begin{align}
\overline{S}(t) &\equiv \sum_{m=-\infty}^{\infty} (m-n_0)^2~ \overline{P}_{m}(t) = \frac{\Delta^2}{\lambda^2} \bigg[ 1 - \ee^{-\lambda t} (1+\lambda t)  \bigg]  +  ({\cal N}-n_0)^2 (1-\ee^{-\lambda t}) , \label{eq:MSD-lam-non0}
\end{align}
which in the limit $\lambda \to 0$ reproduces correctly the result~\eqref{eq:MSD-lam0}. 

Figure~\ref{fig:meas} shows the behavior of average site occupation probability $\overline{P}_m(t)$ as a function of $t$ for two different values of $m$. In the figure, the dashed line in black depicts the analytical result given in Eq.~\eqref{eq:result-rho}, while the continuous line in red is obtained from numerical implementation of the reset dynamics averaged over $4\times 10^3$ realizations. Appendix \ref{app2} provides details on the numerical implementation procedure.  One may observe a good agreement between numerics and analytical results. 

It is evident from Fig.~\ref{fig:meas} that the quantity $\overline{P}_m(t)$ relaxes to a stationary value at long times.  In order to obtain the latter value analytically, we consider Eq.~(\ref{eq:result-rho}) in the limit $t\to \infty$.  In this limit, the first term on the right hand side drops out; using the fact that $J_{m}^2(x)=J_{-m}^2(x)= J_{|m|}^2{(x)}$ and the integral identity~\cite{Gradshteyn:2007}
\begin{align}
\int_{0}^{\infty} \! \! \! {\rm d}x~\ee^{-\alpha x} \,J_\nu(\beta x) J_\nu(\gamma x) = \frac{1}{\pi\sqrt{\gamma \beta} }~ Q_{\nu - \frac{1}{2}}\!\Bigg(  \frac{\alpha^2 + \beta^2 +\gamma^2}{2\beta \gamma}\Bigg)
\end{align}
with the conditions ${\mathrm{Re}}(\alpha+\im \beta +\im \gamma) > 0, ~~\gamma > 0, ~~{\mathrm{Re}}(\nu) > -{1}/{2} $, where $Q_{\nu}(x)$ being the Legendre function of the second kind, we obtain from Eq.~\eqref{eq:result-rho} that the stationary-state distribution is given by
\begin{align}
\overline{P}^{\hskip1.5pt\mathrm{st}}_m &\equiv \lim_{t \to \infty} \overline{P}_m(t) 
= \lambda \int_0^{\infty}  {\rm d}t'~\ee^{-\lambda t'}J_{|m-{\cal N}|}^2(\Delta t') 
= \frac{\lambda}{\pi \Delta}~ Q_{|m-{\cal N}|-\frac{1}{2} } \Bigg(1 + \frac{\lambda^2}{2\Delta^2} \Bigg) . \label{eq:Pst-LFSK}
\end{align}
The function $Q_{\nu}(x)$ may be constructed from the associated Legendre function of the second kind $Q_{\nu}^{\mu}(x)$ with $\mu = 0$. Then, using the relation~\cite[Eqs.~14.3.3 and~14.3.7]{DLMF-NIST}
\begin{align}
Q_{\nu}^{\mu}(x) &= \ee^{\im \mu \pi }~\frac{\sqrt{\pi}~ (x^2-1)^{\mu/2}}{2^{\nu + 1} x^{\nu + \mu +1}} ~\frac{\Gamma(\nu+\mu+1)}{\Gamma(\nu+3/2)} {}_2F_1\bigg(  \frac{\nu}{2} + \frac{\mu}{2} +1, \frac{\nu}{2} + \frac{\mu}{2} + \frac{1}{2} ; \nu+\frac{3}{2}; \frac{1}{x^2} \bigg) ;  \nonumber \\[1ex]
&\hskip70pt \mu+\nu \neq -1,-2,-3, \ldots,
\end{align}
where $\Gamma (z)$ and ${}_2F_1(a,b;c;z)$ are the Gamma function and the Gauss' hypergeometric function, respectively, Eq.~\eqref{eq:Pst-LFSK} may be rewritten as
\begin{align}
\displaystyle \overline{P}^{\hskip1.5pt\mathrm{st}}_m 
&= \frac{\Big[1+ \frac{2\Delta^2}{\lambda^2} \Big]^{-1/2} }{ \Big[2+ \frac{\lambda^2}{\Delta^2} \Big]^{|m-{\cal N}|}} ~\frac{\Gamma\big(|m-{\cal N}|+1/2\big)}{\sqrt{\pi} ~\Gamma\big(|m-{\cal N}|+1\big)}~\nonumber \\
&\hskip50pt \times {}_2F_1\bigg(\frac{|m-{\cal N}|}{2} +\frac{3}{4}, \frac{|m-{\cal N}|}{2} +\frac{1}{4};|m-{\cal N}|+1; \frac{4}{(2+(\lambda^2/\Delta^2))^2} \bigg). \label{eq:Pst-long-time}
\end{align}
In Fig.~\ref{fig:meas}, the stationary values $\overline{P}^{\hskip1.5pt\mathrm{st}}_m$ for the two given values of $m$ are shown by the dash-dotted lines in grey.   We observe a very good agreement between numerics and analytical results. 

In the stationary state, the mean displacement and the MSD of the particle about the initial location $n_0$, denoted by $\overline{\mu}^{\hskip1.5pt\mathrm{st}}$ and $\overline{S}^{\hskip1.5pt\mathrm{st}}$, respectively, are obtained from Eqs.~\eqref{eq:MEAN-lam-non0} and~\eqref{eq:MSD-lam-non0} by taking the limit $t \to \infty$.  One obtains
\begin{align}
\overline{\mu}^{\hskip1.5pt\mathrm{st}} &= {\cal N} - n_0 ,  \label{eq:MEAN-lam-non0-st} \\
\overline{S}^{\hskip1.5pt\mathrm{st}} &= \frac{\Delta^2}{\lambda^2} + ({\cal N} - n_0)^2. \label{eq:MSD-lam-non0-st}
\end{align}

On the basis of the foregoing,  we see a very interesting effect being induced by classical stochasticity introduced through resets at random times.  A particle starting from a site and evolving under the Hamiltonian (\ref{eq:TBC-Hamiltonian}) interspersed with such resets reaches at long times a stationary state in which the particle has time-independent values for the probability to be on different sites.  Concomitantly,  the MSD of the particle about its initial position attains time-independent values.  These aspects imply localization of the quantum particle on the lattice through introduction of classical stochasticity.  This effect offers a stark contrast when viewed against the results in the absence of any reset,  as discussed in Section~\ref{secIII-a}.  In this case,  the site occupation probability continues to change forever as a function of time see Eq.~(\ref{eq:Pmt-bare-Bessel}), and in particular, the MSD about the initial position changes quadratically in time, Eq.~(\ref{eq:MSD-lam0}).  

\begin{figure}[ht]
\centering
        \includegraphics[scale=1.25]{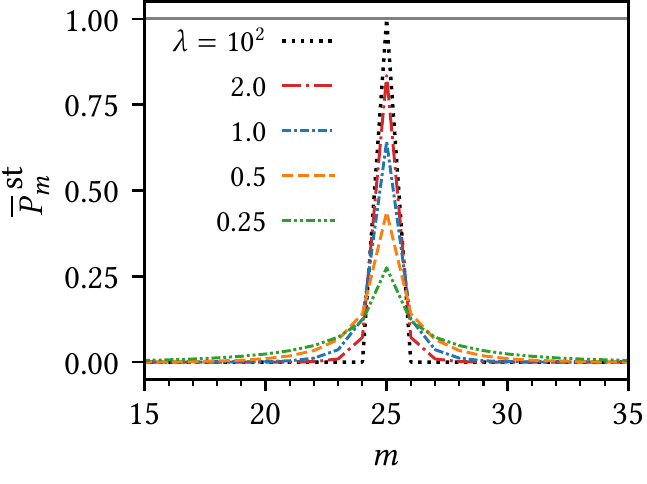}
        \caption{The figure shows for the tight-binding model~\eqref{eq:model-1} subject to repeated resets at random times the steady-state distribution
        $\overline{P}^{\hskip1.5pt\mathrm{st}}_m$ as a function of $m$ for five different values of $\lambda$, namely, $0.25$ (dash-double-dotted line in green), $0.5$ (dashed line in orange), $1.0$ (densely dash-dotted line in blue), $2.0$ (dash-dotted line in red), and $10^2$ (dotted line in black). Here, we have ${\cal N}=25$ and $\Delta =1$. In this plot, the lines are obtained by using Eq.~\eqref{eq:Pst-long-time}. It is seen from the plot that with an increase in $\lambda$ (i.e., with a decrease in time gap between two successive resets) the particle is more likely to be found in the neighbourhood of the detector site ${\cal N}$ in the steady state. The plot suggests that in the limit $\lambda \to \infty$, the particle in the steady state indeed stays at the detector site with probability one, see Eq.~\eqref{eq:Pst-lambda-infinity}.}
        \label{fig:ss}
\end{figure}

It is interesting to evaluate the stationary-state distribution $\overline{P}^{\hskip1.5pt\mathrm{st}}_m$ in the limit $(\lambda/\Delta) \to 0$.  To this end, we first note that~\cite[Sec.~14.2(ii)]{DLMF-NIST} $Q_{\nu}(x) = \Gamma(\nu + 1) \, {\bm{Q}}_{\nu}(x)$, where ${\bm{Q}}_{\nu}(x)$ is the Olver's associated Legendre function of order zero. In the limit $x \to 1^{+}$, one may write~\cite[Eqs.~14.8.9 and~5.2.2]{DLMF-NIST}
\begin{align}
{\bm{Q}}_{\nu}(x) &= -\frac{\ln(x-1)}{2\Gamma(\nu+1)} + \frac{ \ln 2 -2\gamma - 2\Gamma'(\nu+1)/\Gamma(\nu+1)}{2\Gamma(\nu+1)}  + {\cal{O}}(x-1) ;  \quad \nu \neq -1,-2,-3,\ldots , \label{eq:Olvers-ALF}
\end{align}
where we have $\Gamma'(x) \equiv {\mathrm d \Gamma(x)}/{\mathrm d}x$, and $\gamma$ is the Euler-Mascheroni constant, defined as $\gamma \equiv \lim_{n \to \infty} \big( \, \sum_{k=1}^{n} 1/k - \ln \,n \, \big)$. In our case, since we have $\lambda>0$ and $\Delta>0$, the limit $(\lambda/\Delta) \to 0$ reduces to the limit  $(\lambda/\Delta) \to 0^{+}$. Then, from Eqs.~\eqref{eq:Pst-LFSK} and~\eqref{eq:Olvers-ALF}, we obtain
\begin{align}
\lim_{(\lambda/\Delta) \to 0} ~ \overline{P}^{\hskip1.5pt\mathrm{st}}_m  &= \frac{1}{\pi} \bigg(\lim_{x \to \, 0^{+}} x \bigg) \bigg( \lim_{x \to 1^{+}} Q_{|m-{\cal{N}}|-\frac{1}{2}}(x) \bigg) = -\frac{1}{2\pi} \bigg(\lim_{x \to \, 0^{+}} x \bigg) \bigg( \lim_{x \to 1^{+}} \ln(x-1) \bigg) \nonumber \\
&= -\frac{1}{2\pi}~ \lim_{x \to \, 0^{+}} ~x \ln \, x = 0^{+}.
\end{align}
We thus see that in the limit $(\lambda/\Delta) \to 0$, the stationary site occupation probability satisfies $\overline{P}^{\hskip1.5pt\mathrm{st}}_m \to 0~\forall~m$. However, since the probability is normalized, what happens in this case is that the particle spreads out over larger number of lattice sites of the infinite chain as $(\lambda/\Delta) \to 0$.  
In Fig.~\ref{fig:ss}, we have plotted $\overline{P}^{\hskip1.5pt\mathrm{st}}_m$ as a function of $m$ for different values of $\lambda$ while keeping $\Delta $ fixed. It is evident from the figure that with a decrease in $\lambda$, the quantity $\overline{P}^{\hskip1.5pt\mathrm{st}}_m$ gets distributed over more lattice sites centered at the detector site ${\cal N}$.  What this suggests is that eventually in the limit $(\lambda/ \Delta) \to 0$, the particle spreads over the whole infinite lattice chain, yielding $\overline{P}^{\hskip1.5pt\mathrm{st}}_m \to 0$ at a fixed $m$ but nevertheless satisfying  $\sum_{m=-\infty}^\infty \overline{P}^{\hskip1.5pt\mathrm{st}}_m=1$.   Consistently,  the MSD of the particle diverges in this limit,  as may be seen from Eq.~\eqref{eq:MSD-lam-non0-st}.  In the simultaneous limits $(\lambda/\Delta) \to 0$ and $|m-\mathcal{N} | \to \infty\,$, while keeping the product $\lambda |m - \mathcal{N}|/\Delta$ fixed, one obtains by considering the integral in Eq.~\eqref{eq:Pst-LFSK} that $\overline{P}^{\hskip1.5pt\mathrm{st}}_m \approx (1/\pi) (\lambda / \Delta) K_0(\lambda |m-\mathcal{N}|/\Delta)$, where $K_0$ is the zeroth-order modified Bessel function of the second kind~\cite[Appendix B]{qreset1}. 

We now study the limit $(\lambda/\Delta) \to \infty$. 
From the definition of the Gauss' hypergeometric function~\cite{DLMF-NIST}
${}_2F_1(a,b;c;z) = ({\Gamma(c)}/{(\Gamma(a)\Gamma(b))})\sum_{s=0}^{\infty} \,  z^{s} ~{\Gamma(a+s) \Gamma(b+s)}/{(\Gamma(c+s) ~s!)} $,
it is seen that $\lim_{z \to 0} ~{}_2F_1(a,b;c;z) = 1$.  Using this result,  we obtain from Eq.~\eqref{eq:Pst-long-time} that
\begin{align}
\lim_{(\lambda/\Delta) \to \infty}~ \overline{P}^{\hskip1.5pt\mathrm{st}}_m = \delta_{m{\cal N}}. \label{eq:Pst-lambda-infinity}
\end{align}
In the limit $(\lambda/\Delta) \to \infty$,  we get from Eqs.~\eqref{eq:MEAN-lam-non0-st} and~\eqref{eq:MSD-lam-non0-st} the results $\overline{\mu}^{\hskip1.5pt\mathrm{st}} = ({\cal N} - n_0)$ and $\overline{S}^{\hskip1.5pt\mathrm{st}} = ({\cal N} - n_0)^2$, which are consistent with Eq.~\eqref{eq:Pst-lambda-infinity}.
From Fig.~\ref{fig:ss}, it is seen that with the increase of $\lambda$ at a fixed $\Delta$, the distribution $\overline{P}^{\hskip1.5pt\mathrm{st}}_m$ becomes more peaked around $m={\cal N}$, suggesting that in the limit $(\lambda/\Delta) \to \infty$, we would have $\overline{P}^{\hskip1.5pt\mathrm{st}}_m = \delta_{m{\cal N}}$, thereby validating Eq.~\eqref{eq:Pst-lambda-infinity}. 

\subsection{The model subject to projective measurements at random times}
\label{secIII-c}

We next consider the form of $T$ that implements instantaneous projection to a detector state corresponding to a detector placed on an arbitrary site ${\cal N}$.  Specifically,  the dynamics is as follows: The quantum particle while starting at time $t = 0$ from an arbitrary site $n_0$ of the TBC streams unitarily under the Hamiltonian (\ref{eq:TBC-Hamiltonian}) until a random time instant $t_1$ at which an instantaneous projection takes place of the state vector onto the state $|{\cal N}\ra$ corresponding to an arbitrary site ${\cal N}$.  
Following the measurement postulate of quantum mechanics~\cite{Cohen},  a projection of the instantaneous state of a quantum
system onto the Hilbert space of
a measuring device is considered akin to a measurement having been performed on the system by the device,  and in this sense, we may think of the site ${\cal N}$ as containing a detector that performs a measurement on the system.  We will consequently refer to the site ${\cal N}$ as the detector site.  Denoting by $|\psi(t_1\ra_-$ the state just prior to the projection, the state just after the projection is given by $|\psi(t_1)\ra_+=P|\psi(t_1)\ra_-$, with $P \equiv |{\cal N}\ra \la {\cal N}|$ being the projection operator.  Correspondingly, the density operator evolves as $\rho_+(t_1)=P\rho_-(t_1)P^\dagger$.  Note that here unlike the case of reset considered in the preceding subsection, one has only a \textit{partial} collapse of the instantaneous state vector onto the state $|{\cal N}\ra$; the amount of collapse depends on the overlap of the state $|\psi(t_1)\ra_-$  with $|{\cal N}\ra$.  Subsequent to the first projection,  evolution proceeds with the projected state $\rho_-(t_1)$ as an unitary evolution until another random time instant $t_2\, (t_2 > t_1)$,  when a second projection onto the detector site takes place, and so on.  Finally,  we may ask at time $t$ by which a random number of projections have taken place about the probability $P_m(t)$ for the particle to be on an arbitrary site $m$ of the lattice.  Note that $\sum_{m=-\infty}^\infty P_m(t)$ gives the trace of the density operator at time $t$, and that every time there is a measurement, the magnitude of the trace decreases. 

Quantum measurement is an important topic of contemporary interest in the context of measurement theory as well as quantum information. Unlike classical measurement processes, quantum measurements are characterized by non-negligible and irreversible interaction between the measuring apparatus and the system of interest.  The measurement postulate of quantum mechanics asserts that every measurement involves instantaneous projection of the state of the system onto the quantum state dictated by the measuring apparatus~\cite{Cohen}. When such measurements are carried out at regular intervals of time, Misra and Sudarshan had shown in a landmark paper that the system gets stuck in a given state in the limit in which the measurements are sufficiently frequent. This is called the quantum Zeno effect \cite{Misra:1977}.  Over the years, the Zeno effect has been studied in a variety of contexts, e.g., in studying its link with quantum measurement theory \cite{Presilla:1996}, Zeno crossovers in fermionic wires~\cite{Marino1} and spin chains~\cite{Marino2}, bath-induced Zeno localization in the context of driven quantum systems~\cite{Maimbourg:2021}, etc.

Experimentally, as it turns out, it is more convenient to implement a random or a stochastic sequence of measurement protocol~\cite{Gherardini:2016,Muller:2017,Gherardini:2017,Gherardini:2018,Do:2019} rather than the regular set of measurements as envisaged in the original paper of Misra and Sudarshan~\cite{Misra:1977}.  Indeed,  experiments aiming to implement projective measurements to demonstrate the Zeno effect would typically employ a timer to time the gap between successive measurements, and because the timer would invariably be of finite precision that it would be difficult, if not impossible, to ensure that measurements are performed at exactly regular time intervals.  In the case of independent and identically distributed intervals of time between consecutive measurements, it has been analytically demonstrated that the survival probability of the system to remain in the projected state has a large deviation (exponentially decaying) form in the limit of an infinite number of measurements~\cite{Gherardini:2016}. The large deviation form allows to estimate the typical value of the survival probability. This offers the possibility to tune the survival probability by manipulating the probability distribution of the random time intervals between consecutive measurements. When applied to Zeno-protected entangled states,  it is seen that measurements at random times enhance the survival probability when the Zeno limit is not reached~\cite{Gherardini:2016}.  In Ref.  \cite{Muller:2017}, it was shown that by implementing
a strong system-environment interaction switched on
at random times, the dynamics of the system remains confined to a subspace.  Reference \cite{Gherardini:2017} addressed the issue of the validity of ergodicity assumption (that the time average of an observable equals its ensemble average) in the case of an open quantum system that is being continuously perturbed by the environment through performing observations on the system at random times.  In Ref.~\cite{Gherardini:2018},  the authors studied the effect of stochastic fluctuations on the distribution of energy exchanged between a quantum system and the external environment through measurements performed on the system at random times.  It was revealed that the set-up obeys a stochastic version of nonequilibrium fluctuation relations (e.g., the quantum Jarzynski equality). The work serves as a demonstration that the fluctuation relations are robust against presence of randomness in the time intervals between measurements.  Reference~\cite{Do:2019} showed how to combine the controlled manipulation of a quantum two-level system with a sequence of projective measurements in order to have direct access to noise correlation functions.  The effectiveness of the proposed method was tested on a Bose–Einstein condensate of ${}^{87}$Rb atoms realized on an atom chip.  The aforementioned variety of studies invoking measurements at random times motivate the study undertaken in this section.  

The TBM (and related systems) when subject to projective measurements has in recent years been extensively studied in the context of detection problem corresponding to a quantum particle evolving under the dynamics to arrive at a chosen set of sites
\cite{Dhar:2015,Dhar:2015-1,Friedman:2017,Friedman:2017-1,Thiel:2018,Thiel:2019,Lahiri:2019,Meidan:2019,Yin:2019,Thiel:2020,Thiel:2020-1,Dubey:2021,Thiel:2021,Liu:2021,Kessler:2021}.  In the context of the present work,  we briefly summarize the contribution of Ref.~\cite{Kessler:2021} that also considers projective measurements at random times.  However, there is a fundamental difference in the way measurements are implemented in our work and in this reference (that builds on an earlier related work, Ref.~\cite{Varbanov:2008}) and in Ref.~\cite{Shushin:2011} that also considers measurements at random times albeit for a two-level system.  To understand this, let us note that while starting from time $t=0$ and undergoing unitary evolution for time $t_1$, we end up with the density operator just prior to the first measurement as $\rho_-(t_1)$. Subsequent to the first measurement, implemented by the projection operator $P$,  we consider subsequent evolution to be continued with the projected component, so that the density operator just after the first measurement is $\rho_+(t_1)=P\rho_-(t_1)P^\dagger$.  Continuing it this way, the projection operator just after the second measurement at time $t_2$ will be $\rho_+(t_2)=P \rho_-(t_2)P^\dagger$, with $\rho_-(t_2)$ obtained by evolving $\rho_+(t_1)$ unitarily for time $(t_2-t_1)$, and so on.  In contrast to this approach,  Ref.~\cite{Kessler:2021} considers subsequent evolution after every measurement to be continued with the leftover component, namely, what remains of the density operator after a measurement has been performed.  More precisely,  the density operator just after the first measurement is $\rho_+(t_1)=\widetilde{P}\rho_-(t_1)\widetilde{P}^\dagger$, just after the second measurement is $\rho_+(t_2)=\widetilde{P}\rho_-(t_2)\widetilde{P}^\dagger$, and so on, with $\widetilde{P}=\mathbb{I}-P$ and $\mathbb{I}$ the identity operator.  Thus,  our work differs from Refs.~\cite{Shushin:2011,Kessler:2021,Varbanov:2008} in the very nature of the dynamical set-up that is implemented in the two sets of work.  Our work is an addition to this existing literature on the TBM, wherein we adduce a theoretical framework new to this area of research,  in the form of the stochastic Liouville equation that allows to address the quantum dynamics of unitary evolution interspersed at random times with non-unitary interactions modelled by a transition superoperator.  

In the present case,  the form of $T$ should be such that subsequent to a measurement at any time instant $t$, the density operator becomes
$\rho_+(t)=T\rho_-(t)=P\rho_-(t)P^\dagger$, with $P = \ket{{\cal N}}\bra{{\cal N}}$ being the projection operator.  It is easy to see that this is achieved with the form 
\begin{align}
(n_1 n_1'|T|n_2 n_2')=\delta_{n_1 {\cal N}} \delta_{n_1' {\cal N}}\delta_{n_2 {\cal N}}\delta_{n_2' {\cal N}}.
\label{eq:Tmatrix-again}
\end{align}

Referring to the preceding subsection, we see that to study the case at hand, Eq.~(\ref{eq:Tmatrix}) has to be replaced with Eq.~(\ref{eq:Tmatrix-again}), while Eqs.~(\ref{eq:rho0-TBC}),~(\ref{eq:U-identity-0}),~(\ref{eq:U-identity-3}),~(\ref{eq:Pmt-equation}),~(\ref{eq:Pms}),~(\ref{eq:P0s-1}),~(\ref{eq:P0-2}),  (\ref{eq:U0s-result}) continue to hold in the present case.  Hence, $\overline{P}_m^{(0)}(t)$ is given by Eq.~(\ref{eq:P0-2}). Using Eq.~(\ref{eq:Pms}),  we next have
\begin{align}
\widetilde{\overline{P}}_m^{(1)}(s)
&=\lambda\sum_{\substack{n_1,n_2,n_3,\\ n_4,n_5,n_6}}
(mm|\widetilde{U}_0(s)|n_1n_2)(n_1n_2|T|n_3n_4)  (n_3n_4|\widetilde{U}_0(s)|n_5n_6)\la
n_5|\rho(0)|n_6\ra \nonumber \\[1ex]
&=\lambda (mm|\widetilde{U}_0(s)|{\cal N}{\cal N})({\cal N}{\cal N}|\widetilde{U}_0(s)|n_0n_0),
\end{align}
where in obtaining the second equality, we have used Eq.~(\ref{eq:Tmatrix-again}). It then follows on using Eq.~(\ref{eq:U0s-result}) that
\begin{align}
\overline{P}_m^{(1)}(t)&=\lambda~\int_0^t {\rm d}t_1~\left[\ee^{-\lambda (t-t_1)}J^2_{m-{\cal N}}(\Delta (t-t_1))\right]\left[\ee^{-\lambda t_1}J^2_{{\cal N}-n_0}(\Delta t_1)\right] \nonumber \\
&=\lambda~\ee^{-\lambda t}\int_0^t {\rm d}t_1~J^2_{m-{\cal N}}(\Delta (t-t_1))J^2_{{\cal N}-n_0}(\Delta t_1).
\end{align}
Proceeding in the same manner as above, we get
\begin{align}
\widetilde{\overline{P}}_m^{(2)}(s)
&=\lambda^2 (mm|\widetilde{U}_0(s)|{\cal N}{\cal N})({\cal N}{\cal N}|\widetilde{U}_0(s)|{\cal N}{\cal N})({\cal N}{\cal N}|\widetilde{U}_0(s)|n_0n_0),
\end{align}
and hence that
\begin{align}
\overline{P}_m^{(2)}(t)
&=\lambda^2~\int_0^t {\rm d}t_2 \int_0^{t_2} {\rm d}t_1~\left[\ee^{-\lambda (t-t_2)}J^2_{m-{\cal N}}(\Delta (t-t_2))\right]\left[\ee^{-\lambda(t_2-t_1)}J^2_{{\cal N}-{\cal N}}(\Delta (t_2-t_1))\right]\nonumber \\
&\hskip87.6pt \times\left[\ee^{-\lambda t_1}J^2_{{\cal N}-n_0}(\Delta t_1)\right] \nonumber \\
&=\lambda^2~\ee^{-\lambda t}\int_0^t {\rm d}t_2 \int_0^{t_2} {\rm d}t_1~J^2_{m-{\cal N}}(\Delta (t-t_2))J^2_{0}(\Delta (t_2-t_1))J^2_{{\cal N}-n_0}(\Delta t_1),
\end{align}
and in general that for $p\in[1,\infty)$, we have
\begin{align}
\overline{P}_m^{(p)}(t)&=\lambda^p~\int_0^t {\rm d}t_p \int_0^{t_{p}} {\rm d}t_{p-1}\ldots \int_0^{t_3}{\rm d}t_2 \int_0^{t_2}{\rm d}t_1~\left[\ee^{-\lambda (t-t_p)}J^2_{m-{\cal N}}(\Delta (t-t_p))\right]\nonumber \\
&\hskip50pt\times\left[\ee^{-\lambda(t_p-t_{p-1})}J^2_{{\cal N}-{\cal N}}(\Delta (t_p-t_{p-1}))\right]\ldots\left[\ee^{-\lambda (t_2-t_1)}J^2_{{\cal N}-{\cal N}}(\Delta (t_2- t_1))\right]\nonumber \\
&\hskip50pt\times \left[\ee^{-\lambda t_1}J^2_{{\cal N}-n_0}(\Delta t_1)\right] \nonumber \\
&=\lambda^p~\ee^{-\lambda t}\int_0^t {\rm d}t_p \int_0^{t_{p}} {\rm d}t_{p-1}\ldots \int_0^{t_3}{\rm d}t_2 \int_0^{t_2}{\rm d}t_1~J^2_{m-{\cal N}}(\Delta (t-t_p))J^2_{0}(\Delta (t_p-t_{p-1})) \ldots \nonumber \\
&\hskip50pt\times J^2_{0}(\Delta (t_3- t_2))J^2_{0}(\Delta (t_2- t_1))J^2_{{\cal N}-n_0}(\Delta t_1),
\label{eq:Pmp-projective-result}
\end{align}
which when substituted in the expansion $\overline{P}_m(t)=\sum_{p=0}^\infty \overline{P}_m^{(p)}(t)$ yields $\overline{P}_m(t)$.

Equation~(\ref{eq:Pmp-projective-result}) may be interpreted as follows.  Dynamical evolution involving $p\ge 1$ projective measurements comprises unitary evolution, with no measurement, of the particle from the initial site $n_0$ for time $t_1$ (the probability of which is $\exp({-\lambda t_1})$), at which instant a projection happens onto the detector site. Subsequently,  there is unitary evolution, with no measurement, of the particle from the detector site ${\cal N}$ for time $t_2-t_1$ (the probability of which is $\exp[{-\lambda (t_2-t_1)}]$), at which instant a second projection happens, and so on.  The dynamics ends with the last projection at time $t_p$, and unitary evolution, with no measurement, of the particle from the detector site to site of interest $m$ for time $t-t_p$. Integrating over all possible values of $t_1,t_2,\ldots,t_p$ then yields $\overline{P}_m^{(p)}(t)$ in Eq.~(\ref{eq:Pmp-projective-result}). 

\begin{figure*}
\centering
\includegraphics[scale=1.25]{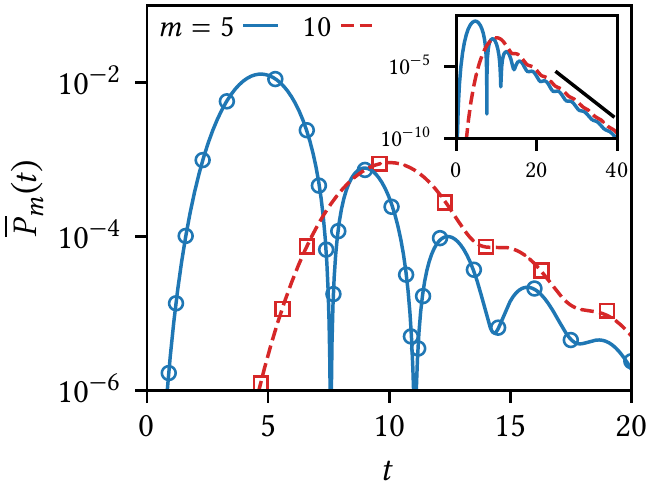}
        \caption{The figure shows for the tight-binding model~\eqref{eq:model-1} subject to projective measurements at random times the dependence of 
        the average site occupation probability 
        $\overline{P}_m(t)$ on $t$ for two values of $m$, namely, $m=5$ and $10$. The initial location of the particle is $n_0 =1$, while the location of the detector is ${\cal N}=10$; also, we have $ \Delta = 1$, and 
        $\lambda=0.5$. In the plot, the lines depict the analytical result computed using Eqs.~\eqref{eq:P0-2} and~\eqref{eq:Pmp-projective-result}, retaining terms up to ${\mathcal{O}}(\lambda^{10})$; we have checked that the higher-order terms do not contribute appreciably to the result. The points are obtained from numerical implementation of the dynamics and involves averaging over $10^4$ realizations and a periodic lattice of $N=30$ sites. The plot shows a very good agreement between the theoretical and the numerical result. The inset shows the exponential decay of $\overline{P}_m(t)$ at long times: $\overline{P}_m(t) \sim \exp(-\lambda t)$.}
        \label{fig:meas-proj}
\end{figure*}

Figure~\ref{fig:meas-proj} shows the behavior of average site occupation probability $\overline{P}_m(t)$ as a function of $t$ for two different values of $m$. In the plot, the lines depict the analytical result computed using Eqs.~\eqref{eq:P0-2} and~\eqref{eq:Pmp-projective-result} by retaining terms up to ${\mathcal{O}}(\lambda^{10})$. The points are obtained from numerical implementation of the dynamics, by averaging over $10^4$ realizations. The plot shows a very good agreement between the theoretical and the numerical result.  In the inset, we show the exponential decay of $\overline{P}_m(t)$ at long times: $\overline{P}_m(t) \sim \exp(-\lambda t)$, implying thereby that $\overline{P}_m(t \to \infty) \to 0$. This is in stark contrast to the case of resetting considered in Sec.~\ref{secIII-b}, where $\overline{P}_m(t)$ relaxes as $t \to \infty$ to a nonzero value, see Fig.~\ref{fig:meas}.

\begin{figure*}
\centering
\includegraphics[scale=1.25]{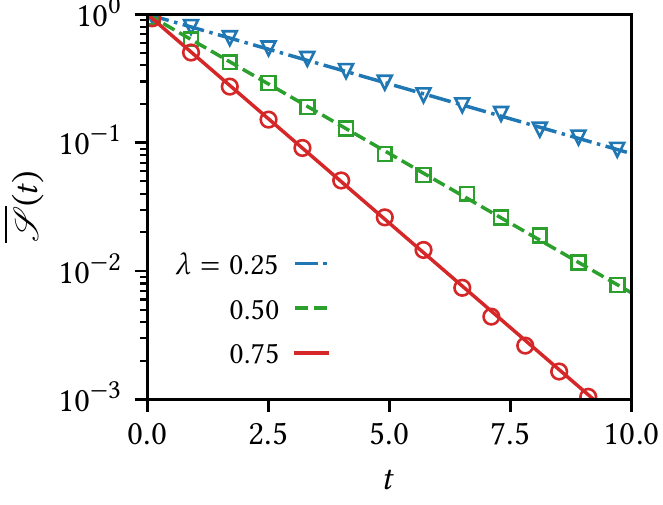}
        \caption{The figure shows for the tight-binding model~\eqref{eq:model-1} subject to projective measurements at random times the dependence of 
        the average survival probability 
        $\overline{\mathcal{S}}(t)$ on $t$ for three different values of $\lambda$, namely, $\lambda = 0.25$ (dash-dotted line in blue), 0.50 (dashed line in green), and 0.75 (solid line in red). The initial location of the particle is $n_0 =1$, while the location of the detector is ${\cal N}=10$; also, we have $ \Delta = 1$. In the plot, the lines depict the behavior $\exp(-\lambda t)$, while the points are obtained from numerical implementation of the dynamics and involves averaging over $5 \times 10^3$ realizations and a periodic lattice of $N=30$ sites. }
        \label{fig:meas-proj-sp}
\end{figure*}

As discussed earlier,  for every realization of the measurement sequence,  the quantity $\sum_{m=-\infty}^\infty P_m(t)$ should decrease as a function of time.  One may define ${\cal S}(t) \equiv \sum_{m=-\infty}^\infty P_m(t)$ as the survival probability that the particle has ``survived'' in the system despite the projective  measurements that reduce the trace of the density operator. The trace is nothing but a measure of  probability for the particle to be found anywhere on the TBC lattice.  Note that the survival probability is a random variable whose value varies from one realization of the measurement sequence to another, and the average quantity $\overline{P}_m(t)$ yields the average survival probability $\overline{\cal S}(t)$. Figure~\ref{fig:meas-proj-sp} shows that the average survival probability behaves as $\overline{\cal S}(t) \sim \exp(-\lambda t)$. One may contrast this exponential decay with the case of reset dynamics considered in the preceding subsection in which one has in the light of Eq.~(\ref{eq:result-rho}) that $\overline{\cal S}(t)=1~\forall~t$. 

The usual Zeno result \cite{Misra:1977}, see also \cite{Venugopalan:2007}, when adapted to our setting would imply that on performing projective measurements to the initial state (i.e., with ${\cal N}=n_0$) at regular intervals of length $\tau \to 0$ (the limit of frequent-enough measurements) over a fixed and finite time $t$,  one has the probability for the particle to be found in the initial state as $P_{n_0}(t) \approx 1- ({\rm No~of~measurements~in~time~}t)({\rm Var}(H))\tau^2 = 1-t\tau({\rm Var}(H))$; here, ${\rm Var}(H)$ is the variance of energy in the initial state $|n_0\rangle$.  Thus, to leading order,  one has the Zeno result that $P_{n_0}(t)=1$. With the variance ${\rm Var}(H)$ equal to $\Delta^2/2$, as may be straightforwardly calculated by using the representation (\ref{eq:model-1}),  we finally have $P_{n_0}(t) \approx 1-t\tau \Delta^2/2$. 

To discuss the Zeno limit in our case,  we need to consider the limit $\lambda \to \infty$ at a fixed and finite $\Delta$.  This may be seen by noting that $1/\lambda$ is the average time between two measurements,  and hence,  for fixed and finite $t$,  if we want measurements to be done frequently enough on the average, we need to consider the case of $\lambda \to \infty$ at a fixed and finite $\Delta$. To study this limit,  let us rewrite Eq.~(\ref{eq:Pmp-projective-result}) with $m=n_0={\cal N}$ as
\begin{align}
\overline{P}_{n_0}^{(p\ge 1)}(t)&=\ee^{-\lambda t}\int_0^{\lambda t} {\rm d}\tau_p \int_0^{\tau_{p}} {\rm d}\tau_{p-1}\ldots \int_0^{\tau_3}{\rm d}\tau_2 \int_0^{\tau_2}{\rm d}\tau_1~J^2_{0}((\Delta/\lambda) (\lambda t-\tau_p))  \nonumber \\
&\hskip10pt \times  J^2_{0}((\Delta/\lambda) (\tau_p-\tau_{p-1})) \ldots  J^2_{0}((\Delta/\lambda) (\tau_3- \tau_2))J^2_{0}((\Delta/\lambda)(\tau_2- \tau_1))J^2_{0}((\Delta/\lambda) \tau_1).
\label{eq:Pmp-projective-result-zeno}
\end{align}
We want to evaluate the right hand side in the limit $\lambda \to \infty$ at a fixed and finite $\Delta$.  To this end, noting that $J_{0}(x) = 1 - ({x^2}/{4})$ for small $x$~\cite[Eq.~10.2.2]{DLMF-NIST}, implying $J^2_0(x) \approx 1-x^2/2$, we get
\begin{align}
\overline{P}_{n_0}^{(p\ge 1)}(t)& \approx \ee^{-\lambda t}\int_0^{\lambda t} {\rm d}\tau_p \int_0^{\tau_{p}} {\rm d}\tau_{p-1}\ldots \int_0^{\tau_3}{\rm d}\tau_2 \int_0^{\tau_2}{\rm d}\tau_1 \nonumber \\
&\hskip20pt\times \left[1-\frac{1}{2}\left(\frac{\Delta}{\lambda}\right)^2 \left\{ (\lambda t-\tau_p)^2+(\tau_p-\tau_{p-1})^2+\ldots+(\tau_3-\tau_2)^2+(\tau_2-\tau_1)^2 \right\} \right]\nonumber \\
&=\ee^{-\lambda t}\frac{(\lambda t)^p}{p!}-\frac{\ee^{-\lambda t}}{2}\left(\frac{\Delta}{\lambda}\right)^2 \int_0^{\lambda t} {\rm d}\tau_p \int_0^{\tau_{p}} {\rm d}\tau_{p-1}\ldots \int_0^{\tau_3}{\rm d}\tau_2 \int_0^{\tau_2}{\rm d}\tau_1\nonumber \\
&\hskip20pt\times \left[(\lambda t-\tau_p)^2+(\tau_p-\tau_{p-1})^2+\ldots+(\tau_3-\tau_2)^2+(\tau_2-\tau_1)^2+\tau_1^2\right] \nonumber \\
&= \ee^{-\lambda t} (\lambda t)^p \left[ \frac{1}{p!} - \Delta^2 t^2 \frac{p+1}{(p+2)!} \right] \, .
\label{eq:Pmp-projective-result-zeno-1}
\end{align}
Using Eqs.~(\ref{eq:P0-2}) and~\eqref{eq:Pmp-projective-result-zeno-1}, in the limit $\lambda \to \infty$ at a fixed and finite $\Delta$, we then obtain
\begin{align}
\overline{P}_{n_0}(t)& \approx \ee^{-\lambda t} J^2_{0}(\Delta t) + \sum_{p=1}^\infty \ee^{-\lambda t} (\lambda t)^p \left[ \frac{1}{p!} - \Delta^2 t^2 \frac{p+1}{(p+2)!} \right]\nonumber \\
&=1-\frac{\Delta^2t}{\lambda} + \frac{\Delta^2}{\lambda^2} + \ee^{-\lambda t}\left[J^2_{0}(\Delta t)-1+ \frac{\Delta^2 t^2}{2}-\frac{\Delta^2}{\lambda^2}\right] \nonumber \\
&\approx 1-\frac{\Delta^2t}{\lambda} \, .
\label{eq:Zeno-exponential}
\end{align}
We may contrast the above result with the result $P_{n_0}(t) \approx 1-t\tau \Delta^2/2$ for measurements done at regular intervals.  On performing measurements at random intervals distributed according to an exponential,  one may naively replace $\tau$ by its average given by $1/\lambda$ in the above result.  The expression so obtained matches with the result~(\ref{eq:Zeno-exponential}) except for a factor of $2$ in the denominator of the second term. 
We thus see that the amount of suppression in the temporal decay of the probability at the initial location on performing measurements to the initial state at random times is comparable to the one in conventional Zeno effect scenarios, which however does not require performing measurements at exactly regular intervals.  As we have discussed in the initial part of this section,  performing measurements at random times is much more amenable to experimental realization than those at regular intervals. 

\begin{figure*}
\centering
\includegraphics[scale=1.25]{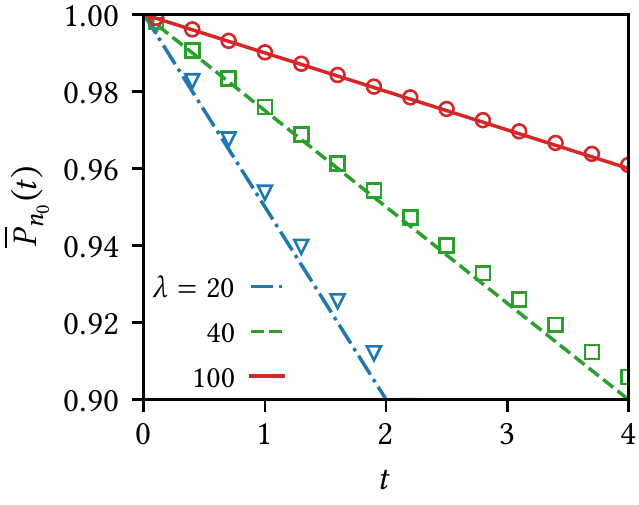}
        \caption{The figure shows for the tight-binding model~\eqref{eq:model-1} subject to projective measurements at random times the Zeno limit with $\Delta =1 $ as $\lambda \to \infty$. In the plot, the lines depict the analytical result given in Eq.~\eqref{eq:Zeno-exponential}, while the points are obtained from numerical implementation of the dynamics and involving averaging over $4\times 10^3$ realizations and a periodic lattice of $N=30$ sites with $m = n_0 = {\cal N}=10$. }
        \label{fig:zeno}
\end{figure*}

Figure~\ref{fig:zeno} shows the validity of the analytical result~\eqref{eq:Zeno-exponential}. In the plot, while the lines depict the analytical result, the points are obtained from numerical implementation of the dynamics and involving averaging over $4\times 10^3$ realizations and a periodic lattice of $N=30$ sites with $m = n_0 = {\cal N}=10$. We see that indeed the analytical result~\eqref{eq:Zeno-exponential} holds for large $\lambda$.

\section{Conclusions}
\label{secIV}

What happens when a generic quantum system evolving unitarily in time is subject to repeated instantaneous interactions with the environment, modelled as non-unitary interventions, at random times? In this work, we developed a general theoretical framework to address this issue and in particular to obtain the average density operator of the system at any time during the dynamical evolution.  The formalism applies to any quantum system, to any form of interaction, and to any form of the distribution of the random times between successive interactions that we considered to be independent and identically-distributed random variables.  In order to demonstrate an explicit application of the formalism, we  addressed within the ambit of a representative quantum system, the so-called tight-binding chain (TBC), the physical effects ensuing from performing resets and quantum projective measurements on the system at exponentially-distributed random times.  Our work serves as an illustration of the effectiveness of the density operator description \textit{a la} the SLE in handling stochastic averages over a background classical probability distribution while dealing with quantum unitary evolution interrupted at random times with non-unitary interventions. 

While our analysis is developed here for the case of the tight-binding chain, our general technique would apply to any quantum system. It would thus be of great interest to extend our analysis to other quantum systems, and in particular, to the case of the tight-binding chain in presence of a time-varying external field and subject to resets and projective measurements. Here, an intricate interplay of the frequency of variation of the field with the average time between successive interactions is expected to lead to interesting dynamical effects. 

A further direction of great interest is to consider the probability distribution of the random time gap between successive interactions to be given by a power-law or a L\'{e}vy distribution, instead of the exponential distribution considered in this work. While all moments of an exponential distribution are finite, not so is the case with a power-law distribution for which in specific range of values of the power or the exponent, the mean and the variance of the distribution are infinite. This latter fact is expected to have important bearings on the dynamics of the particle. Investigations in these directions are under way and will be reported elsewhere. 

Another direction worth pursuing is to study in the case of stochastic resetting the issue of relaxation to the stationary state.  Similar studies in the classical setting lead to many interesting phenomena including dynamical transitions~\cite{Majumdar:2015}, and hence it would be worthwhile to contrast such scenarios with what happens in the context of quantum dynamics.

\section{Acknowledgements}

SG acknowledges support from the Science and Engineering Research
Board (SERB), India under SERB-TARE scheme Grant No.
TAR/2018/000023, SERB-MATRICS scheme Grant No.
MTR/2019/000560, and SERB-CRG scheme Grant No. CRG/2020/000596. SG also thanks ICTP–Abdus Salam International Centre for Theoretical Physics, Trieste, Italy, for support under its Regular Associateship scheme.  SD is grateful to the Indian National Science Academy for support through their Senior Scientist Scheme.  This work was carried out using computational facilities of the Advanced Computing Research Centre,
University of Bristol, UK - \texttt{http://www.bristol.ac.uk/acrc/}. 
\section{Appendix A: Liouville operators and the notion of a superoperator}
\label{app1}
\setcounter{equation}{0}

Following Ref. \cite{Dattagupta:1987}, the Liouville operator ${\cal L}$ associated with an ordinary operator such as the Hamiltonian $H$ and when operating on an ordinary operator $A$ gives rise to the commutator of $H$ with $A$:
\begin{align}
{\cal L}\,A=[ \, H,\, A \,].
\label{eq:app1}
\end{align}
Thus, a Liouville operator operating on an ordinary operator gives another ordinary operator, just as a wave function operated on by an operator yields another wave function, and in this sense,  a Liouville operator is a superoperator.  Hence, all the rules of linear algebra are valid in the operations with superoperators.  Accordingly, we denote the `states’ of a Liouville operator by round kets:
\begin{align}
{\cal L}|nm)={\cal L}|n\rangle \langle m|=H|n\rangle \langle m|-|n\rangle \langle m|H,
\end{align}
where the complete set of states are in our case the Wannier states, namely, the set $\{|n\rangle\}$.                                         
The superoperator then lives in a product Hilbert space, the dimension of which is the square of the space of the associated operator.  We see that the `matrix elements' of the superoperator are labelled by four indices,  just as those of an ordinary  operator are labelled by two indices.  One has the closure property for the two-indexed states as $\sum_{\,m,\,n}|mn)(mn|=\mathbb{I}$.
Consistent with Eq.~(\ref{eq:app1}) then,  we have
\begin{align}
\langle n|{\cal L}A|m\rangle=\sum_{m',n'}(nm|{\cal L}|n'm')\langle n'|A|m'\rangle,
\end{align}
wherein 
\begin{align}
(nm|{\cal L}|n'm')=\langle n |H|n'\rangle \delta_{mm'}-\langle m'|H|m\rangle \delta_{nn'}.
\label{eq:matrix-elements-app}
\end{align}
Finally, the Heisenberg time-evolution of an operator may be defined as
\begin{align}
A(t)=\ee^{{\rm i}{\cal L}t}A(0)=\ee^{{\rm i}Ht}A(0)\ee^{-{\rm i}Ht},
\end{align} 
while that of the density operator is given by
\begin{align}
\rho(t)=\ee^{-{\rm i}{\cal L}t}\rho(0)=\ee^{-\im Ht}\rho(0)\ee^{\im Ht}.
\label{eq:liouville-eqn-app}
\end{align}

The rule~(\ref{eq:matrix-elements-app}) applies to any superoperator,  and applying it to the density operator $\rho(t)$ given in Eq.~(\ref{eq:liouville-eqn-app}), we get 
\begin{align}
\langle m |\rho(t)|n\rangle&=\sum_{m', \, n'}(mn|\ee^{-{\rm i}{\cal L}t}|m'n')\langle m'|\rho(0)|n'\rangle 
=\sum_{m',\, n'}\langle m|\ee^{-{\rm i}Ht}|m'\rangle\langle m'|\rho(0)|n'\rangle \langle n'|\ee^{{\rm i}Ht}|n\rangle.
\label{eq:rho-matrix-element}
\end{align}
Comparing the first and the second equality in Eq.~(\ref{eq:rho-matrix-element}) yields the matrix elements of 
the superoperator $\exp({-\im \mathcal{L}t})$ as
\begin{align}
( mn| \ee^{-\im \mathcal{L}t} | m'n' ) = \la m | \ee^{-\im H t} |m'\ra \la n' | \ee^{\im H t} |n\ra . \label{eq:iLt-elements}
\end{align}

\section{Appendix B: Numerical implementation of the dynamical protocol introduced in the main text for the TBC Hamiltonian}
\label{app2}

Let us first show how starting from a one-dimensional periodic chain of $N$
sites, the model~(\ref{eq:model-1}) arises in the limit $N \to \infty$.
We start with the Hamiltonian
\begin{align}
H=-\frac{\Delta}{2}({K}+{K}^\dagger); \quad {K}=\sum_{n=-N/2}^{N/2-1}
\ket{n}\bra{n+1},
\label{eq:model-2}
\end{align}
with 
\begin{align}
\ket{q}=\sum_{n=-N/2}^{N/2-1} \ee^{-\im 2\pi qn/N}\ket{n}, \quad
\ket{n}=\frac{1}{N}\sum_{q=-N/2}^{N/2-1} \ee^{\im 2\pi
qn/N}\ket{q}. \label{eq:DFT-00}
\end{align}
In writing the above equations, we have considered $N$ to be even. For odd $N$, one has $n, q \in [-(N-1)/2,(N-1)/2]$. Now, in Eq.~\eqref{eq:DFT-00}, effecting the change of variable $q \to k\equiv
2\pi q/N$, so that $\Delta k=2\pi/N \to 0$ as $N \to \infty$, we get in
the limit $N \to \infty$ that
\begin{align}
\begin{split}
\ket{k}=\sum_{n=-\infty}^\infty \ee^{-\im kn}\ket{n}, \quad
\ket{n}=\frac{1}{N}\frac{N}{2\pi}\sum_{k=-\pi }^{\pi -2\pi/N} \Delta k~
\ee^{\im kn}\ket{k} \stackrel{N \to \infty}{\to} 
\frac{1}{2\pi}\int_{-\pi}^\pi {\rm d}k~\ee^{\im kn}\ket{k}.
\end{split}
\end{align}
Redefining the Fourier transforms as 
\begin{equation}
\begin{aligned}
\ket{k}=\frac{1}{\sqrt{2\pi}}\sum_{n=-\infty}^\infty \ee^{-\im kn}\ket{n}, \quad
\ket{n}=\frac{1}{\sqrt{2\pi}}\int_{-\pi}^\pi {\rm d}k~\ee^{\im kn}\ket{k}, \label{eq:DFT} 
\end{aligned}
\end{equation}
we see that Eq.~(\ref{eq:model-2}) in the limit $N \to \infty$ and
together with Eq.~(\ref{eq:DFT}) are equivalent to the model defined in
Section~\ref{secII}. We therefore expect numerical evolution using the TBC for a large-enough value of $N$ to generate results that would match with theoretical results obtained in the main text in the limit $N \to \infty$.

In our numerical implementation of the dynamical protocol,  we employ a one-dimensional periodic lattice of finite number of sites equal to $N$.  The density operator (and any other ordinary operator,  e.g., the Hamiltonian $H$) is then given by an $N \times N$ matrix, while the interaction superoperator $T$ (and any other superoperator) is represented by a matrix of dimension $N^2 \times N^2$.  A particular realization of the dynamical protocol over a total duration $t$ involves sampling first the time gaps $\tau$ between successive interactions with the external environment. Here, we sample the interval $\tau$'s independently from the exponential distribution~(\ref{eq:ptau-exponential}) using standard techniques, and record their sum after each draw until the sum exceeds the value of $t$. For example, let us assume that after ${\mathfrak N}$-th draw, we have $\sum_{p=1}^{\mathfrak N} \tau_p < t$, and the very next draw yields such a $\tau_{{\mathfrak N}+1}$ that one gets $\sum_{p=1}^{{\mathfrak N}+1} \tau_p > t$. At this point, as the total time of evolution of the system is given to be $t$, we stop the iteration of sampling the $\tau$'s and set the value of the last interval as $\tau_{{\mathfrak N}+1} = t-\sum_{p=1}^{\mathfrak N} \tau_p$. Following this procedure, for a single realization of the dynamical protocol, we end up with the sequence $\{ \tau_1, \tau_2, \ldots, \tau_{\mathfrak N}, \tau_{{\mathfrak N}+1} = t -\sum_{p=1}^{\mathfrak N} \tau_p\}$ of $({\mathfrak N}+1)$ elements. The first ${\mathfrak N}$ elements of the sequence are sampled from the exponential distribution, while the sum of all the elements yields $t$.   
Note that the number ${\mathfrak N}$ varies from one realization to another.  The dynamical evolution proceeds as follows: We start with the initial density operator $\rho(0)$,  which has only its $(n_0,n_0)$-th element nonzero and in fact equal to unity, with $n_0$ being the initial location of the particle. We evolve $\rho(0)$ unitarily in time for time $\tau_1$ to obtain the evolved density operator as $\rho(\tau_1)=\exp(-{\rm i}H\tau_1)\rho(0)\exp({\rm i}H\tau_1)$,  and then operate on it by $T$ constructed according to the particular type of interaction with the external environment that we want to investigate. The result is evolved unitarily for time $\tau_2$,  then operated on by $T$, and so on.  The final step of evolution comprises unitary evolution for time $\tau_{{\mathfrak N}+1}$,  and the $(mm)$-th element of the resulting matrix gives the probability for the given realization of the dynamical protocol for the particle to be found on site $m$ at time $t$.  The process is repeated for a large-enough number of dynamical realizations to finally obtain the average site occupation probability $\overline{P}_m(t)$ as reported in the main text.     

\vspace{1cm}



\begin{thebibliography}{99}
\bibitem{Note:Bessel}
Note that on the right hand side of Eqs.~\eqref{eq:Pmt-bare-Bessel} and~\eqref{eq:result-rho}, one would have $- \, \Delta $ in the argument of the Bessel function if one had taken the TBC Hamiltonian to be negative of what we have considered, namely, Eq.~\eqref{eq:TBC-Hamiltonian}. Nevertheless, since $J_{m}^2(x) = J_{m}^2(-x)$, this would leave $P_{n_0}(t)$ and $\overline{P}_{n_0}(t)$ invariant.
\bibitem{note-last-reset} The exponential distribution (\ref{eq:ptau-exponential}) implies that in a small time $0<\Delta t \ll 1$,  the probability of a reset to take place equals $\lambda\Delta t$, while the same for no reset is $1-\lambda\Delta t$.  To see this, let us ask for the probability for the next reset to happen after a certain time $t$, in the interval $[t,t+{\rm d}t]$? Let us discretize time in equal steps of length $0 < \Delta t \ll 1$; we will in the end consider the limit $\Delta t \to 0$.  The number of such steps during time duration $t$ equals $n_{\rm d}=t/\Delta t$.   Consequently,  the probability that the next reset takes place after time $t$, in the interval $[t,t+\Delta t]$,  equals $\underbrace{(1-\lambda\Delta t)(1-\lambda\Delta t)\ldots(1-\lambda\Delta t)}_{n_{\rm d}~\mathrm{factors}}\lambda\Delta t=(1-\lambda\Delta t)^{n_{\rm d}}\lambda\Delta t$.  In the limit $\Delta t \to 0$, the latter quantity equals $\exp(-\lambda t)\,\lambda  {\rm d}t$.  Thus, the probability for no reset to occur during a given time duration $t$ equals $\exp(-\lambda t)$, while the probability for the next reset to occur after time $t$, that is, in the interval $[t,t+{\rm d}t]$, equals $\exp(-\lambda t)\,\lambda{\rm d}t$.  It then follows that at a given time instant $t$,  the probability that the last reset happened in the interval $[t-\tau-{\rm d}\tau, t-\tau]$, with $\tau \in [0,t]$, equals $\exp(-\lambda\tau)\,\lambda{\rm d}\tau$.
\end{thebibliography}
\end{document}